\documentstyle[11pt]{article}

\input amssym
\catcode`\@=11\relax
\newwrite\@unused
\def\typeout#1{{\let\protect\string\immediate\write\@unused{#1}}}

\def\psglobal#1{
\immediate\special{ps: plotfile #1 }}
\def\psfiginit{\typeout{psfiginit}
\immediate\psglobal{figtex.pro}%
\special{ps:: /TeXMagnification {\the\mag} def}%marco:look @ figtex.pro.
}

\def\@nnil{\@nil}
\def\@empty{}
\def\@psdonoop#1\@@#2#3{}
\def\@psdo#1:=#2\do#3{\edef\@psdotmp{#2}\ifx\@psdotmp\@empty \else
    \expandafter\@psdoloop#2,\@nil,\@nil\@@#1{#3}\fi}
\def\@psdoloop#1,#2,#3\@@#4#5{\def#4{#1}\ifx #4\@nnil \else
       #5\def#4{#2}\ifx #4\@nnil \else#5\@ipsdoloop #3\@@#4{#5}\fi\fi}
\def\@ipsdoloop#1,#2\@@#3#4{\def#3{#1}\ifx #3\@nnil
       \let\@nextwhile=\@psdonoop \else
      #4\relax\let\@nextwhile=\@ipsdoloop\fi\@nextwhile#2\@@#3{#4}}
\def\@tpsdo#1:=#2\do#3{\xdef\@psdotmp{#2}\ifx\@psdotmp\@empty \else
    \@tpsdoloop#2\@nil\@nil\@@#1{#3}\fi}
\def\@tpsdoloop#1#2\@@#3#4{\def#3{#1}\ifx #3\@nnil
       \let\@nextwhile=\@psdonoop \else
      #4\relax\let\@nextwhile=\@tpsdoloop\fi\@nextwhile#2\@@#3{#4}}
\def\psdraft{
	\def\@psdraft{0}
	\def\@psdraftspecial{100}
}
\def\psdraftspecial{
	\def\@psdraft{0}
	\def\@psdraftspecial{0}
}
\def\psfull{
	\def\@psdraft{100}
}
\psfull

\newif\if@prologfile
\newif\if@postlogfile
\newif\if@bbllx
\newif\if@bblly
\newif\if@bburx
\newif\if@bbury
\newif\if@height
\newif\if@width
\newif\if@rheight
\newif\if@rwidth
\newif\if@clip
\newif\if@right
\newif\if@left
\newif\if@toplines
\newif\if@box
\newif\if@caption
\newif\if@surround
\newif\if@captionwidth
\newif\if@captionwrite
\newif\if@captionopen
\def\@p@@sclip#1{\@cliptrue}
\def\@p@@sfile#1{%\typeout{file is #1}
		\def\@p@sfile{#1}
}
\def\@p@@sfigure#1{
		\def\@p@sfile{#1}
}
\def\@p@sfake{\hbox to 0pt{\hss Whatever\hss}}
\def\@p@@sbbllx#1{
		\@bbllxtrue
		\@d@mscratch=#1
		\edef\@p@sbbllx{\number\@d@mscratch}
}
\def\@p@@sbblly#1{
		\@bbllytrue
		\@d@mscratch=#1
		\edef\@p@sbblly{\number\@d@mscratch}
}
\def\@p@@sbburx#1{
		\@bburxtrue
		\@d@mscratch=#1
		\edef\@p@sbburx{\number\@d@mscratch}
}
\def\@p@@sbbury#1{
		\@bburytrue
		\@d@mscratch=#1
		\edef\@p@sbbury{\number\@d@mscratch}
}
\def\@p@@sheight#1{
		\@heighttrue
		\@d@mscratch=#1
   		\edef\@p@sheight{\number\@d@mscratch}
}
\def\@p@@swidth#1{
		\@widthtrue
		\@d@mscratch=#1
		\edef\@p@swidth{\number\@d@mscratch}
}
\def\@p@@srheight#1{
		\@rheighttrue
		\@d@mscratch=#1
		\edef\@p@srheight{\number\@d@mscratch}
}
\def\@p@@srwidth#1{
		\@rwidthtrue
		\@d@mscratch=#1
		\edef\@p@srwidth{\number\@d@mscratch}
}
\def\@p@@sright#1{\@righttrue \@surroundtrue}
\def\@p@@sleft#1{\@lefttrue \@surroundtrue}
\def\@p@@sextraheight#1{\@d@mextraheight=#1}
\def\@p@@sbox#1{\@boxtrue}
\def\@p@@scaption#1{\@captiontrue}
\def\@p@@stoplines#1{
		\@toplinestrue
		\@c@ttoplines=#1
}
\def\@p@@scaptionwidth#1{
		\@captionwidthtrue
	  	\@d@mcaptionwidth=#1
}
\def\@p@@scaptionwrite#1{
		\global\@captionwritetrue
		\global\@w@rname=\expandafter{\jobname_captions.tex}
		\typeout{Captions are written to \the\@w@rname}
}

\def\@p@@sprolog#1{\@prologfiletrue\def\@prologfileval{#1}}
\def\@p@@spostlog#1{\@postlogfiletrue\def\@postlogfileval{#1}}
\def\@cs@name#1{\csname #1\endcsname}
\def\@setparms#1=#2,{\@cs@name{@p@@s#1}{#2}}
\def\ps@init@parms{
		\@bbllxfalse \@bbllyfalse
		\@bburxfalse \@bburyfalse
		\@heightfalse \@widthfalse
		\@rheightfalse \@rwidthfalse
		\def\@p@sbbllx{}\def\@p@sbblly{}
		\def\@p@sbburx{}\def\@p@sbbury{}
		\def\@p@sheight{}\def\@p@swidth{}
		\def\@p@srheight{}\def\@p@srwidth{}
		\def\@p@sfile{}
		\def\@p@scost{10}
		\def\@sc{}
		\@prologfilefalse
		\@postlogfilefalse
		\@clipfalse
		\@rightfalse \@leftfalse
		\@boxfalse \@captionfalse
		\@toplinesfalse \@surroundfalse
		\@d@mextraheight=0pt
 		\@c@ttoplines=0
		\@pshape={} \def\@p@srheight@total{}
		\@captionwidthfalse \@d@mcaptionwidth=0pt
}
\def\parse@ps@parms#1{
	 	\@psdo\@psfiga:=#1\do
		   {\expandafter\@setparms\@psfiga,}}
\newif\ifno@bb
\newif\ifnot@eof
\newread\ps@stream
\newtoks\@linetok
\def\bb@missing{
	\typeout{psfig: searching \@p@sfile \space  for bounding box}
	\openin\ps@stream=\@p@sfile
	\no@bbtrue
	\not@eoftrue
	\catcode`\%=12
	\loop
		\read\ps@stream to \line@in
		\global\@linetok=\expandafter{\line@in}
		\ifeof\ps@stream \not@eoffalse \fi
		\@bbtest{\@linetok}
		\if@bbmatch\not@eoffalse\expandafter\bb@cull\the\@linetok\fi
	\ifnot@eof \repeat
	\catcode`\%=14
}	
\catcode`\%=12
\newif\if@bbmatch
\def\@bbtest#1{\expandafter\@a@\the#1%%BoundingBox:\@bbtest\@a@}
\long\def\@a@#1%%BoundingBox:#2#3\@a@{
     \ifx\@bbtest#2\@bbmatchfalse\else\@bbmatchtrue\fi}
\long\def\bb@cull#1 #2 #3 #4 #5 {
	\@d@mscratch=#2 bp\edef\@p@sbbllx{\number\@d@mscratch}
	\@d@mscratch=#3 bp\edef\@p@sbblly{\number\@d@mscratch}
	\@d@mscratch=#4 bp\edef\@p@sbburx{\number\@d@mscratch}
	\@d@mscratch=#5 bp\edef\@p@sbbury{\number\@d@mscratch}
	\no@bbfalse
}
\catcode`\%=14
\def\compute@bb{
		\no@bbfalse
		\if@bbllx \else \no@bbtrue \fi
		\if@bblly \else \no@bbtrue \fi
		\if@bburx \else \no@bbtrue \fi
		\if@bbury \else \no@bbtrue \fi
		\ifno@bb \bb@missing \fi
		\ifno@bb \typeout{FATAL ERROR: no bb supplied or found}
			\no-bb-error
		\fi
		\count203=\@p@sbburx
		\count204=\@p@sbbury
		\advance\count203 by -\@p@sbbllx
		\advance\count204 by -\@p@sbblly
		\edef\@bbw{\number\count203}
		\edef\@bbh{\number\count204}
}
\def\in@hundreds#1#2#3{\count240=#2 \count241=#3
		     \count100=\count240	% 100 is first digit #2/#3
		     \divide\count100 by \count241
		     \count101=\count100
		     \multiply\count101 by \count241
		     \advance\count240 by -\count101
		     \multiply\count240 by 10
		     \count101=\count240	%101 is second digit of #2/#3
		     \divide\count101 by \count241
		     \count102=\count101
		     \multiply\count102 by \count241
		     \advance\count240 by -\count102
		     \multiply\count240 by 10
		     \count102=\count240	% 102 is the third digit
		     \divide\count102 by \count241
		     \count200=#1\count205=0
		     \count201=\count200
			\multiply\count201 by \count100
		     	\advance\count205 by \count201
		     \count201=\count200
			\divide\count201 by 10
		     	\multiply\count201 by \count101
			\advance\count205 by \count201
		     \count201=\count200
			\divide\count201 by 100
			\multiply\count201 by \count102
			\advance\count205 by \count201
		     \edef\@result{\number\count205}
}
\def\compute@wfromh{
		\in@hundreds{\@p@sheight}{\@bbw}{\@bbh}
		\edef\@p@swidth{\@result}
}
\def\compute@hfromw{
		\in@hundreds{\@p@swidth}{\@bbh}{\@bbw}
		\edef\@p@sheight{\@result}
}
\def\compute@handw{
		\if@height
			\if@width
			\else
				\compute@wfromh
			\fi
		\else
			\if@width
				\compute@hfromw
			\else
				\edef\@p@sheight{\@bbh}
				\edef\@p@swidth{\@bbw}
			\fi
		\fi
}
\def\compute@resv{
		\if@rheight \else \edef\@p@srheight{\@p@sheight} \fi
		\if@rwidth \else \edef\@p@srwidth{\@p@swidth} \fi
		\edef\@p@srheight@total{\@p@srheight}
}
\newtoks\@pshape
\def\@c@ttoplines{\count120}
\def\@c@theightcount{\count121}
\def\@c@tshapecount{\count122}
\newdimen\@d@mwidthshape
\newdimen\@d@mextraheight
\newdimen\@d@mscratch
\def\compute@parshape{
	\if@right
		\if@left
	   		\typeout{error: Can't have both left and right set}
			\@leftfalse
		\fi
	\fi
	\@d@mscratch=\@p@swidth truesp
	\divide \@d@mscratch by 19
	\multiply \@d@mscratch by 20
	\edef\@p@swidthdimen{\the\@d@mscratch}
	\@c@tshapecount=\@c@ttoplines
 	\@d@mscratch=\baselineskip
	\multiply \@d@mscratch by \@c@ttoplines
	\advance \@d@mscratch by .4\baselineskip
    	\edef\@p@stopdistance{\the\@d@mscratch }
	\@d@mscratch=\@p@sheight truesp
	\divide \@d@mscratch by 2
	\edef\@p@shalfboxheight{\the\@d@mscratch}
	\if@toplines
		\loop \@pshape=\expandafter{\the\@pshape 0pt \hsize}
		\advance\@c@ttoplines by -1
		\ifnum\@c@ttoplines>0 \repeat
	\fi
   	\@c@theightcount=\@p@srheight@total
	\advance \@c@theightcount by \@d@mextraheight
	\divide  \@c@theightcount by \baselineskip
	\advance \@c@theightcount by 1
    	\advance \@c@tshapecount by \@c@theightcount
	\advance \@c@theightcount by -1
	\@d@mwidthshape=\hsize
     	\advance \@d@mwidthshape by -\@p@swidthdimen
	\if@left
		\def\@moveshape{0pt}
		\ifnum\@c@theightcount>0
		      	\loop
			\@pshape=%
			\expandafter{\the\@pshape %
					\@p@swidthdimen \@d@mwidthshape}
			\advance \@c@theightcount by -1
			\ifnum\@c@theightcount>0 \repeat
		\else
			\advance \@c@tshapecount by 1
		\fi
	\fi
	\if@right
		\@d@mscratch=\hsize
		\advance \@d@mscratch by -\@p@swidth truesp
		\edef\@moveshape{\@d@mscratch}
		\ifnum\@c@theightcount>0
			\loop
			\@pshape=\expandafter{\the\@pshape 0pt \@d@mwidthshape}
			\advance \@c@theightcount by -1
			\ifnum\@c@theightcount>0 \repeat
		\else
			\advance \@c@tshapecount by 1
		\fi
	\fi
	\ifnum \@p@srheight=0
		\@pshape={}
		\@c@tshapecount = 0
	\else
	 	\@pshape=\expandafter{\the\@pshape 0pt \hsize}
	\fi
}
\def\@p@ssetsurroundboxes{
	\global\parshape=\count122 \the\@pshape		% \@c@tshapecount
 	\moveright\@moveshape
	\vbox to 0pt\bgroup\hskip0pt\vskip\@p@stopdistance
}
\newtoks\@captiontok
\newbox\@b@xcaption
\newdimen\@d@mcaptionwidth
\newdimen\@d@mcaptionheight
\newwrite\@w@rcaption
\newtoks\@w@rname
\def\setcaption#1{\@captiontok={#1}}
\def\@set@caption{
	\setbox\@b@xcaption\vbox{\hsize\@d@mcaptionwidth
	\tolerance=9000 \baselineskip=11.4pt
	\noindent\relax\the\@captiontok}
	\if@captionwrite
		\if@captionopen
		\else
			\global\@captionopentrue
			\immediate\openout\@w@rcaption=\the\@w@rname
		\fi
		\immediate\write\@w@rcaption{\the\@captiontok}
		\immediate\write\@w@rcaption{}
	\fi
}
\def\compute@caption{
	\if@captionwidth
	\else
		\@d@mcaptionwidth = \@p@swidth truesp
		\divide \@d@mcaptionwidth by 20
		\multiply \@d@mcaptionwidth by 17
	\fi
	\@set@caption
	\@d@mcaptionheight=\ht\@b@xcaption
	\if@rheight
	\else
		\count100=\@p@srheight
	   	\advance \count100 by \@d@mcaptionheight
	   	\advance \count100 by \bigskipamount
	   	\advance \count100 by \medskipamount
	   	\edef\@p@srheight@total{\number\count100}
	\fi
}
\newif\if@alreadyjtem \@alreadyjtemfalse
\def\newpar{\hfil\vadjust{\vskip\parskip}%
	{\count100=\parskip
	\count101=\baselineskip
	\divide\count101 by 10  \multiply\count101 by 3
	\advance \count100 by \count101
	\divide\count100 by \baselineskip
	\advance\count100 by \prevgraf
	\global\prevgraf=\count100}%
	\break\if@alreadyjtem\else\indent\fi%
}
\let\sav@par=\par
\def\jtem#1{%
    	\if@alreadyjtem\else\bgroup\fi
	\def\par{\sav@par\egroup\sav@par}
	\if@alreadyjtem\else\leftskip\parindent\fi
	\@alreadyjtemtrue
	\noindent\hskip0pt
	\llap{#1\ }\ignorespaces
}
\def\compute@sizes{%
	\compute@bb
	\compute@handw
  	\compute@resv
	\if@caption
		\compute@caption
	\fi
	\if@surround
		\compute@parshape
	\fi
}
\def\@p@sdospecials{
	\ifnum\@p@scost<\@psdraft
	       	\typeout{psfig: including \@p@sfile \space }
	\fi
	\special{ps::[begin] 	\@p@swidth \space \@p@sheight \space
			\@p@sbbllx \space \@p@sbblly \space
			\@p@sbburx \space \@p@sbbury \space
			startTexFig \space }
	\ifnum\@p@scost<\@psdraft
		\if@clip
			\typeout{(clip)}
			\special{ps:: \@p@sbbllx \space \@p@sbblly \space
				\@p@sbburx \space \@p@sbbury \space
			    	doclip \space }
		\fi
	\fi
	\if@box
		\typeout{(box)}
  		\special{ps:: \@p@sbbllx \space \@p@sbblly \space
			\@p@sbburx \space \@p@sbbury \space
		    	dobox \space }
	\fi
	\ifnum\@p@scost<\@psdraft
		\if@prologfile
	    		\special{ps: plotfile \@prologfileval \space }
		\fi
		\special{ps: plotfile \@p@sfile \space }
    		\if@postlogfile
			\special{ps: plotfile \@postlogfileval \space }
		\fi
	\fi
	\special{ps::[end] endTexFig \space }
}
\newif\if@putinvbox

\def\psfig#1{{%
	\ifhmode%
		\vbox\bgroup
		\@putinvboxtrue
	\else
		\@putinvboxfalse
	\fi
       	\ps@init@parms
	\parse@ps@parms{#1}
       	\compute@sizes
	\if@surround
		\psfig@for@surround
	\else
		\psfig@for@regular
	\fi
	\if@putinvbox
       		\egroup
	\fi
}}
\def\psfig@for@surround{%
	\@p@ssetsurroundboxes
	\ifnum\@p@scost<\@psdraft
		\@p@sdospecials
		\vbox to \@p@srheight true sp{\vss}
       	\else
		\if@box
			\@p@sdospecials
		\fi
		\vbox to \@p@srheight true sp{
			\vskip\@p@shalfboxheight
			\hbox to \@p@srwidth true sp{
				\hss
				\ifnum\@p@scost<\@psdraftspecial
					\@p@sfile
				\else
					\@p@sfake
				\fi
      				\hss
			}
		\vss
		}
	\fi
	\if@caption
		\medskip
		\hbox to \@p@srwidth true sp{
			\hss
			\box\@b@xcaption
			\hss
		}
 		\medskip
	\fi
	\vss\egroup
	\vskip-\parskip
}

\def\psfig@for@regular{%
	\if@putinvbox
	\else
		\vskip\parskip
	\fi
	\ifnum\@p@scost<\@psdraft
		\@p@sdospecials
		\vbox to \@p@srheight true sp{%
			\hbox to \@p@srwidth true sp{
			\hfil
			}
		\vfil
		}
       	\else
		\if@box
			\@p@sdospecials
		\fi
	    	\vbox to \@p@srheight true sp{
			\vss
			\hbox to \@p@srwidth true sp{
				\hss
				\ifnum\@p@scost<\@psdraftspecial
					\@p@sfile
				\else
					\@p@sfake
				\fi
				\hss
			}
		    	\vss
		}
	\fi
	\if@caption
		\medskip
		\hbox to \@p@srwidth true sp{
			\hss
			\box\@b@xcaption
			\hss
		}
		\bigskip
	\fi
	\if@putinvbox
	\else
		\vskip-\parskip
	\fi
}
\catcode`\@=12\relax
  
\psfiginit

\font\scriptsizebbfont=msbm7 scaled \magstep 1
\font\footnotesizebbfont=msbm9 scaled \magstep 0
\font\smallbbfont=msbm7 scaled \magstep 2
\font\bbfont=msbm9 scaled \magstep1  % 1.2 pt
 1
 2
 3
 4

\def\scriptsizeBbb#1{\hbox{\scriptsizebbfont #1}}

\def\footnotesizeBbb#1{\hbox{\footnotesizebbfont #1}}
\def\smallBbb#1{\hbox{\smallbbfont #1}}
\def\Bbb#1{\hbox{\bbfont #1}}

\newcommand{\Aff}{\mbox{\it Aff}\,}
\newcommand{\Aut}{\mbox{\it Aut}\,}
\newcommand{\Cone}{\mbox{\it Cone}\,}
\newcommand{\CP}{{\Bbb C}{\rm P}}
\newcommand{\CS}{\mbox{\it CS}\,}
\newcommand{\Def}{\mbox{\it Def}\,}
\newcommand{\Dev}{\mbox{\it Dev}\,}
\newcommand{\Diag}{\mbox{\it Diag}\,}
\newcommand{\Diff}{\mbox{\it Diff}\,}
\newcommand{\Id}{\mbox{\it Id}}
\newcommand{\GL}{\mbox{\it GL}\,}
\newcommand{\Hom}{\mbox{\it Hom}\,}
\newcommand{\Isom}{\mbox{\it Isom}\,}
\newcommand{\Li}{\mbox{\it Li}\,}
\newcommand{\MCG}{\mbox{\it MCG}\,}

\newcommand{\PSL}{\mbox{\it PSL}\,}
\newcommand{\SL}{\mbox{\it SL}}
\newcommand{\SO}{\mbox{\it SO}\,}
\newcommand{\Sp}{\mbox{\it Sp}\,}
\newcommand{\Span}{\mbox{\it Span}\,}

\newcommand{\im}{\mbox{{\rm Im}\,}}
\newcommand{\image}{\mbox{\it Im}\,}
\newcommand{\length}{\mbox{\it length}\,}
\newcommand{\lk}{\mbox{\it lk}\,}

\newcommand{\mod}{\mbox{\rm mod}\,}
\newcommand{\pr}{\mbox{\rm pr}}
\newcommand{\re}{\mbox{\rm Re}\,}

\newcommand{\torsion}{\mbox{\it torsion}\,}
\newcommand{\vol}{\mbox{\it vol}\,}

\newcommand{\itGamma}{{\it \Gamma}}
\newcommand{\itSigma}{{\it \Sigma}}

\begin{document}

\enlargethispage{23cm}

\begin{titlepage}

$ $

\vspace{-1.5cm} % Re: -1.5cm for PC; -2.5cm for UT-Math-system

\noindent\hspace{-1cm}
\parbox{6cm}{\footnotesize January 1999 \newline
              Micron MPC P166 by Marco Monti}\
   \hspace{7.5cm}\
   \parbox{5cm}{ {\tt hep-th/9902092}\newline
                      {ut-ma/99001}}

\vspace{1.5cm}     

%title
\centerline{\large\bf
 On K3-Thurston 7-manifolds and their deformation space:}
\vspace{1ex}
\centerline{\large\bf
 A case study with remarks on general K3T
                                 and M-theory compactification}

\vspace{1.5cm}

%author-'n-address
\centerline{\large Chien-Hao Liu\footnotemark}
\vspace{1.1em}
\centerline{\it Department of Mathematics}
\centerline{\it University of Texas at Austin}
\centerline{\it Austin, Texas 78712-1082}
\footnotetext{E-mail: chienliu@math.utexas.edu}

%%authors-'n-addresses
%\begin{center}
%\parbox[t]{5cm}{
% \centerline{\large Jacques Distler\footnotemark}
% \vspace{1.1em}
% \centerline{\it Department of Physics}
% \centerline{\it University of Texas at Austin}
% \centerline{\it Austin, Texas 78712-1081} }
% \footnotetext{E-mail: distler@golem.ph.utexas.edu}  \
%\hspace{2cm} \
%\parbox[t]{5cm}{
% \centerline{\large Chien-Hao Liu\footnotemark}
% \vspace{1.1em}
% \centerline{\it Department of Mathematics}
% \centerline{\it University of Texas at Austin}
% \centerline{\it Austin, Texas 78712-1082}  }
% \footnotetext{E-mail: chienliu@math.utexas.edu}
%\end{center}

\vspace{2em}

%abstract%
\begin{quotation}
\centerline{\bf Abstract}
\vspace{0.3cm}

\baselineskip 12pt  %13pt for [12pt] style
{\small
M-theory suggests the study of $11$-dimensional space-times
compactified on some $7$-manifolds. From its intimate relation to
superstrings, one possible class of such $7$-manifolds are those
that have Calabi-Yau threefolds as boundary. In this article, we
construct a special class of such $7$-manifolds, named as
{\it K3-Thurston} (K3T) $7$-manifolds. The factor from the K3 part
of the deformation space of these K3T $7$-manifolds admits a
K\"{a}hler structure, while the factor of the deformation space
from the Thurston part admits a special K\"{a}hler structure. The
latter rings with the nature of the scalar manifold of a vector
multiplet in an $N=2$ $d=4$ supersymmetric gauge theory. Remarks
and examples on more general K3T $7$-manifolds and issues to
possible interfaces of K3T to M-theory are also discussed.
} %endsmall
\end{quotation}

\bigskip

\baselineskip 12pt


\bigskip

\noindent {\small
MSC number 1991: 55R05, 53C15, 51M10, 58D27, 83E30
} % end-small

\bigskip

\baselineskip 11pt
% Re: 11 pt for [11pt] style; 12pt for [12pt] style

{\footnotesize
\noindent{\bf Acknowledgements.}
We would like to thank
 Orlando Alvarez and William Thurston
for influential educations,
 Philip Candelas
for group meeting,
 Jacques Distler, Daniel Freed, and Alan Reid
for discussions, lectures, and literatures,
 Volker Braun,
 Hung-Wen Chang, Chong-Sun Chou,
 Xenia de la Ossa,
 Rukmini Dey,
 Robert Gompf,
 Mark Haskins, Pei-Ming Ho, Jacques Hurtubise,
 Vadim Kaplunovsky, Nurit Krausz,
 Robert McNee,
 Rafael Nepomechie,
 Lorenzo Sadun, Margaret Symington, 
 Karen Uhlenbeck, 
 and Frederic Zamora
for valuable inspirations, discussions, and correspondences. 
Special thanks to J.D.\ for many discussions on the related
physics at various stages of the work and for his patience in
explaining things to me and to A.R.\ for many hallway discussions
on hyperbolic geometry.
Finally, thanks to Ling-Miao Chou for the indispensable moral
support.
} %endfootnotesize

\noindent
\underline{\hspace{20em}}

$^1${\footnotesize E-mail: chienliu@math.utexas.edu}

\end{titlepage}

%paper
\newpage
$ $

\vspace{-4em}  % Re: -4cm for PC; -6cm for UT-Math-system

%short heading
\centerline{\sc
 K3-Thurston $7$-Manifolds}

\vspace{1em}

\baselineskip 14pt  %Re: 14pt for [11pt] style
                    %Re: 15pt for [12pt] style.

\begin{flushleft}
{\Large\bf 0. Introduction and outline.}
\end{flushleft}

\begin{flushleft}
{\bf Introduction.}
\end{flushleft}
M-theory anticipates the space-time to be $11$-dimensional
compactified on a $7$-dimensional space. A class of such $7$-spaces
that have appeared in the literature on M-theory compactification
are Joyce manifolds ([Ac1, Ac2] and [Joy]). Many of these manifolds
can be realized as K3-fibrations over Euclidean $3$-orbifolds ([Li]).
A natural extension of this from a $3$-dimensional geometor's point
of view is to consider $7$-manifolds that are K3-fibred over a
hyperbolic $3$-orbifolds. We call such $7$-manifolds
{\it K3-Thurston (K3T)} $7$-manifolds. If one also allows these
$7$-manifolds to have Calabi-Yau threefolds as boundary, then one
may allow the base to be hyperbolic $3$-orbifolds with
${\Bbb Z}\oplus{\Bbb Z}$-cusps. They form the set of hyperbolic
$3$-orbifolds of finite volume. In this article, we construct a
very simple kind of K3T $7$-manifolds and study their deformation
space.

To make this paper more accessible to physicists, some essential
background or references are collected in Sec.\ 1. In Sec.\ 2, using
a K3 surface $X$ with antisymplectic involution, representations of
a $3$-manifold group $\pi_1(M^3)$ to ${\Bbb Z}_2$, and the technique
of bundle-filling, we construct a class of K3T $7$-manifolds that
are either closed or with boundary the Calabi-Yau threefolds
$\cup_{i=1}^h {\Bbb T}^2\times X$. Concrete examples of such are
given, with base some hyperbolic link complements in $S^3$. We then
turn to the study of the deformation space $\Def(\mbox{K3T})$ of
this class of K3T $7$-manifolds in Sec.\ 3. We discuss the
K\"{a}hler factor of $\Def(\mbox{\rm K3T})$ from the K3 part of
K3T and the special K\"{a}hler factor of $\Def(\mbox{\rm K3T})$ from
the Thurston part of K3T.
Remarks on general K3T $7$-manifolds, some ingredients to construct
them, and some examples are given in Sec.\ 4. Finally, we discuss in
Sec.\ 5 possible interfaces from K3T to M-theory. The field
theoretical contents when M-theory is compactified on such
$7$-manifolds will await future work.

\bigskip

\noindent
{\bf Convention.}
Since both real and complex manifolds are involved in this article,
to avoid confusion, a {\it real} $n$-dimensional manifold will be
called an {\it $n$-manifold} while a {\it complex} $n$-dimensional
manifold an {\it $n$-fold}.

\bigskip

\begin{flushleft}
{\bf Outline.}
\end{flushleft}
{\footnotesize
\baselineskip 10pt 

\begin{quote}
 1. Essential mathematical background for physicists.

 2. The construction of a class of K3T $7$-manifold.
  \vspace{-1ex}
  \begin{quote}
   \hspace{-1.3em}
   2.1 \parbox[t]{10cm}{The two ingredients:
       K3 surface with involution and hyperbolic \newline
       $3$-manifolds of finite volume.}\\[0ex]

   \hspace{-1.3em}
   2.2 The construction of K3T $7$-manifolds by bundle-filling.

   \hspace{-1.3em}
   2.3 Examples from the link complements in $S^3$.
  \end{quote}

 3. The deformation space.
  \vspace{-1ex}
  \begin{quote}
   \hspace{-1.3em}
   3.1 The deformation space of K3T 7-manifolds constructed.

   \hspace{-1.3em}
   3.2 The K3 part of the deformations of K3T.

   \hspace{-1.3em}
   3.3 The Thurston part of the deformations of K3T.
  \end{quote}

 4. Remarks and examples on general K3T 7-manifolds.
   
 5. Issues on applications to M-theory compactification.
\end{quote}
} %endfootnotesize

\bigskip

\newpage

\section{Essential mathematical background for physicists.}

We collect in this section some essential mathematical background
or references for the convenience of physicists and also for the
introduction of terminologies and notations.
Additional necessary facts are stated in the related sections.

\bigskip

\noindent $\bullet$
{\bf K3 surfaces and their deformation space.}
Physicists are referred to [As] for a very nice exposition.
A standard reference is [B-P-VV].
Degenerations and isolated singularities of a K3 surface are
 referred to [F-S], [Ku]. [Pe], and [P-P].
A-D-E surface singularities and their monodromy diffeomorphism
 are referred to [Di].
See also [A-M1,A-M2], [Dol], and [Sc].

\bigskip

\noindent $\bullet$
{\bf Hyperbolic geometry.}
([B-P], [C-F-K-P], [C-R], [M-T] and [Th1-Th5].)
A {\it hyperbolic $3$-manifold} is a Riemannian $3$-manifold of
 constant negative sectional curvature, usually normalized to $-1$.
 Up to isometry, there is a unique complete simply-connected one:
 the {\it hyperbolic $3$-space ${\Bbb H}^3$}. All other complete
 hyperbolic $3$-manifolds are quotients of ${\Bbb H}^3$. There are
 several analytic models for ${\Bbb H}^3$; two of them are
 particularly important to us:
 \begin{quote}
  \hspace{-1.9em}(1)
  {\it The upper half-space model}$\,$:
  $$
   {\Bbb H}^3={\Bbb C}\times{\Bbb R}_+=\{(z,t)\,|\,z\in{\Bbb C},t>0\}
   \hspace{1em}\mbox{with metric}\hspace{1em}
   ds^2=\frac{dz\cdot d\overline{z}+dt^2}{t^2}\,.
  $$

  \hspace{-1.9em}(2)
  {\it The Poincar\'{e} ball model}$\,$:
  $$
   {\Bbb H}^3=\{(x_1,x_2,x_3)\in{\Bbb R}^3|x_1^2+x_2^2+x_3^2<1\}
   \hspace{1em}\mbox{with metric}\hspace{1em}
   ds^2=\frac{4\,(\,dx_1^2
                +dx_2^2+dx_3^2\,)}{{(\,1-x_1^2-x_2^2-x_3^2\,)}^2}\,.
  $$
 \end{quote}
(Cf.\ {\sc Figure 1-2}(a).)

The group $\Isom^+({\Bbb H}^3)$ of orientation-preserving
 isometries of ${\Bbb H}^3$ is the same as the M\"{o}bius group
 $\PSL(2,{\Bbb C})$ that acts on the {\it ideal boundary} of
 ${\Bbb H}^3$
 $$
  S^2_{\infty}\; =\;\partial_{\infty}{\Bbb H}^3\;
  =\;\CP^1\; =\;{\Bbb C}\cup\{\infty\}\;
  =\;\{(x_1,x_2,x_3)\in{\Bbb R}^3|x_1^2+x_2^2+x_3^2=1\}
 $$
 by linear fractional transformations. Note that $\PSL(2,{\Bbb C})$
 is isomorphic to $\SO^{\uparrow}_+(1,3)$, the identity component
 of the Lorentz group.

All hyperbolic $3$-manifolds discussed in this paper will be assumed
to be orientable.

\bigskip

\noindent $\bullet$
{\bf Developing map and holonomy.} ([Th1, Th5].)
Given a hyperbolic $3$-manifold $M^3$, perhaps with boundary, then
 one can cover $M^3$ with a hyperbolic coordinate charts
 $(U_{\alpha}, \phi_{\alpha}:U_{\alpha}\rightarrow {\Bbb H}^3)$
 so that all $U_{\alpha}, U_{\alpha}\cap U_{\beta}$ are
 topologically a ball. The transition function
 $\phi_{\alpha\beta}=\phi_{\alpha}\circ\phi_{\beta}^{-1}:
   \phi_{\beta}(U_{\alpha}\cap U_{\beta})
              \rightarrow \phi_{\alpha}(U_{\alpha}\cap U_{\beta})$
 then coincides with the restriction to
 $\phi_{\beta}(U_{\alpha}\cap U_{\beta})$ of a unique element,
 also denoted by $\phi_{\alpha\beta}$, in $\Isom({\Bbb H}^3)$.
Fix an initial chart $U_0$ and a point $p_0$ in the interior of
 $U_0$. Let $\gamma:[0,1]\rightarrow M^3$ be a path at $p_0$
 and $(U_0,\phi_0), (U_1,\phi_1)\,\cdots\,,(U_n,\phi_n)$ be a
 sequence of hyperbolic charts along $\gamma$, i.e.\
 $\gamma\subset\cup_{i=0}^n\,U_i$ and $U_i\cap U_{i+1}$ nonempty
 for $i=0,\,\ldots\,,n-1$.
 One can form a new sequene of hyperbolic charts
 $(U_i,\phi_i^{\,\prime})$, $i=0,\,\ldots\,,n$, by setting
 $$
  \phi_0^{\,\prime}\;=\;\phi_0\,, \hspace{2em}
  \phi_1^{\,\prime}\;=\;\phi_{01}\circ\phi_1\,, \hspace{2em}
   \cdots\,, \hspace{2em}
  \phi_n^{\,\prime}\; =\;\phi_{01}\circ\phi_{12}\circ\,
                           \cdots\,\phi_{n-1,n}\circ\phi_n\,,
 $$
 where $\phi_{i,i+1}$ are understood as elements in
 $\Isom({\Bbb H}^3)$.
 A key feature of $(U_i,\phi_i^{\,\prime})$ is that the new
 transition functions $\phi_{i,i+1}^{\,\prime}$ have now become
 the identity map. Thus we shall call the sequence
 $(U_i,\phi_i^{\,\prime})$, $i=0,\,\ldots\,,n$, the
 {\it analytic continuation} of $(U_0,\phi_0)$ along $\gamma$.
 Such continuation depends only on the initial chart $(U_0,\phi_0)$
 and the homotopy class $[\gamma]$ of $\gamma$ relative to its
 end-points and we will denote the germ associated with
 $\phi_n^{\,\prime}$ by $\phi_0^{[\gamma]}$. ({\sc Figure 1-1}.)
 \begin{figure}[htbp]
  \setcaption{{\sc Figure 1-1.}
   \baselineskip 14pt
   The analytic continuaion of hyperbolic patches along a path
   $\gamma$ (cf.\ {\sc Fig.\ 3.15} in [Th5]).
  } % end-setcaption
  \centerline{\psfig{figure=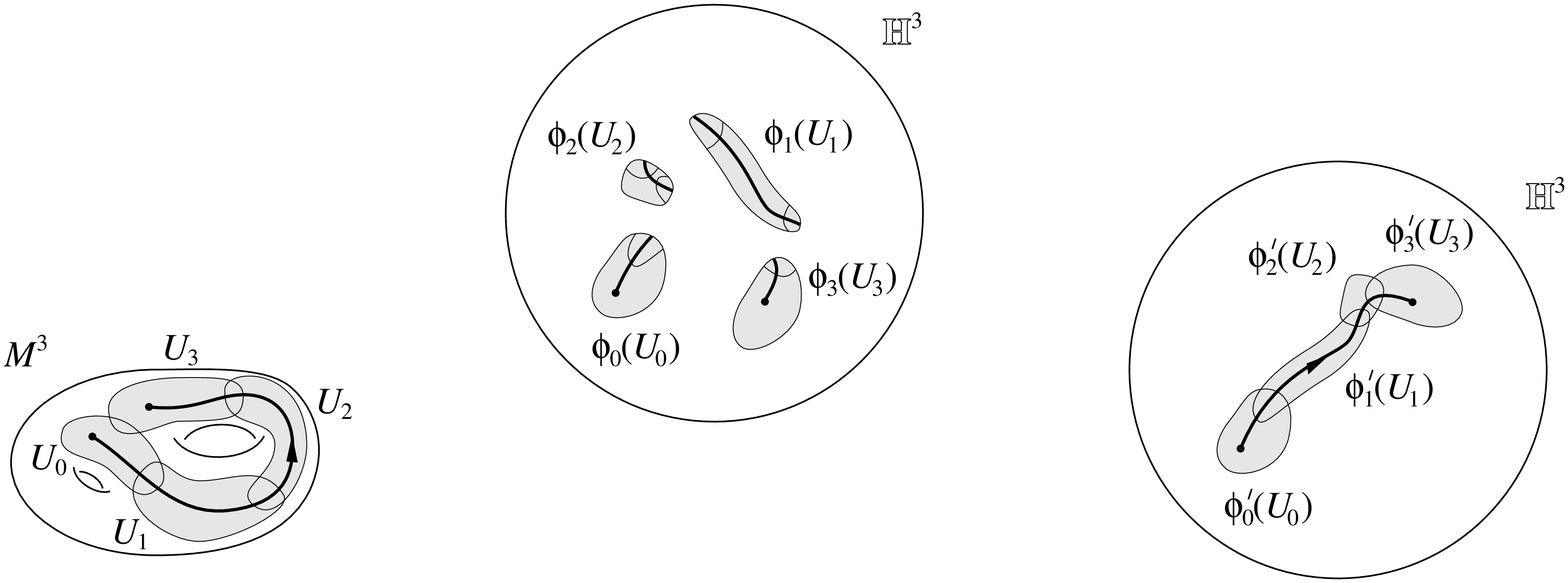,width=13cm,caption=}}
 \end{figure}

Fix a lifting of $U_0$ to the universal covering $\widetilde{M^3}$
 of $M^3$, with the lifted hyperbolic structure, then a local
 isometry
 $$
   \Dev\;:\;\widetilde{M^3}\;\longrightarrow\;{\Bbb H}^3
 $$
 can be uniquely defined via analytic continuation. This map is
 called the {\it developing map} of $M^3$. It depends only on the
 choice of $(U_0,\phi_0)$ and its lifting and, hence, is unique
 up to a post-composition by an isometry of ${\Bbb H}^3$.
Let $\sigma$ be an element of $\pi_1(M^3,p_0)$ and $\alpha$ be its
 representative. Then the {\it holonomy} $g_{\sigma}$ of $\sigma$
 is by definition the unigue $g_{\sigma}\in\Isom({\Bbb H}^3)$ such
 that $\phi_0^{\,\sigma}=g_{\sigma}\circ\phi_0$. 
Let $T_{\sigma}$ be the deck transformation on $\widetilde{M^3}$
 associated to $\sigma$, then
 $\Dev\circ T_{\sigma}=g_{\sigma}\circ \Dev$. From this, one can
 deduce that the map $\mu:\sigma\mapsto g_{\sigma}$ is a group
 homomorphism, called the {\it holonomy homomorphism}. For $M^3$
 oriented, one has
 $$
  \mu\;:\;\pi_1(M^3)\;\longrightarrow\;\PSL(2;{\Bbb C})\,,
 $$
 unique up to the conjugation by an isometry of ${\Bbb H}^3$.

These concepts apply also to general geometric structures
e.g.\ Euclidean, spherical, or affine structures, on a manifold. 

\bigskip

\noindent $\bullet$
{\bf Ideal tetrahedron.} ([Th1] and [Th4].)
An {\it ideal tetrahedron} in ${\Bbb H}^3$ is a
 $3$-simplex $\Delta^3$ in ${\Bbb H}^3$ inscribed in $S^2_{\infty}$
 such that all its faces are totally geodesic. These faces are by
 themselves ideal triangles, as indicated in {\sc Figure 1-2}(a).
 All ideal triangles are isometric to each other. 
Given an edge $e$ of an oriented ideal tetrahedron $\Delta^3$.
 Suppose that $e$ is assigned a temporary orientation, then the two
 faces of $\Delta^3$, with the induced orientation from that of
 $\Delta^3$, that contains $e$ can be distinguished as ``right"
 if the induced orientation on $e$ from that face is the same as
 the assigned orientation of $e$, or ``left" if the induced
 orientation on $e$ from that face is the opposite of the assigned
 orientation of $e$. There is then a unique
 $g\in\Isom^+({\Bbb H}^3)$ that sends the right face to the left
 face while preserving the two end-points of $e$. This defines a
 number $z(e)\in {\Bbb C}$ with $\|z\|$ being the translation
 distance of $g$ along $e$ and $\arg(z)$ being the angle of
 rotation of $g$. If the opposite orientation of $e$ is chosen,
 then the role of left and right for the two faces that contains
 $e$ are reversed. Thus $g$ is replaced by $g^{-1}$, but $z(e)$
 remains the same. The complex number $z(e)$ is thus called
 the {\it edge invariant} of $\Delta^3$ associated to
 (unoriented) $e$. Any of these edge invariants determines the
 ideal tetrahedron up to an isometry. They correspond to the cross
 ratio of the four points of inscription when appropriately ordered.
 We shall denote $\Delta^3$ by $\Delta(z)$, where $z$ is the edge
 invariant of some edge of $\Delta^3$, and call $\Delta(z)$
 {\it the ideal tetrahedron of modulus $z$}.
 ({\sc Figure 1-2} (a) and (b).)

\begin{figure}[htbp]
 \setcaption{{\sc Figure 1-2.}
 \baselineskip 14pt
  In (a), an ideal tetrahedron $\Delta(z)$ of modulus $z$ is shown
   both in the Poincar\'{e} ball and the upper half-space model
   of ${\Bbb H}^3$.
  In (b), the edge invariants associated to $\Delta(z)$ are indicated.
 } % end-setcaption
 \centerline{\psfig{figure=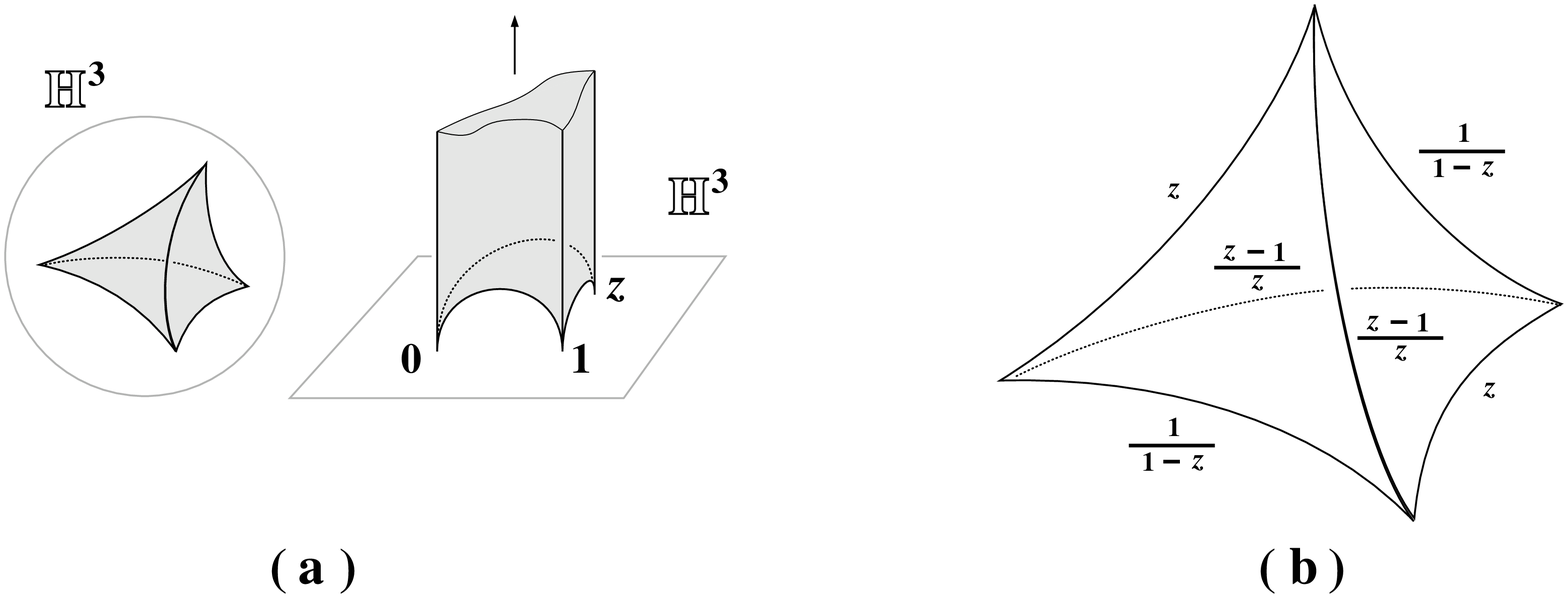,width=13cm,caption=}}
\end{figure}

The importance of understanding ideal tetrahedra first is due to
 the fact proved by Thurston in [Th4] that {\it any hyperbolic
 $3$-manifold $M^3$ of finite volume can be obtained by pasting a
 collection of ideal tetrahedra along their faces}. We shall call
 this an {\it ideal triangulation} of $M^3$.
 Since $M^3$ is assumed to be orientable, one may assume that the
 moduli of these ideal tetrahedra lie in the upper half-plane
 ${\Bbb H}_+=\{z\in{\Bbb C}\,|\,\im(z)>0\}$.
 {\sc Figure 1-3} indicates how an ideal tetrahedron in such
 triangulation may be embedded in $M^3$.

\begin{figure}[htbp]
 \setcaption{{\sc Figure 1-3.}
 \baselineskip 14pt
  How an ideal tetrahedron $\Delta(z)$ in an ideal triangulation
  for a cusped hyperbolic $3$-manifold $M^3$ is indicated.
  Notice that a leg of $\Delta(z)$ may go into an end of $M^3$
  or wind around a simple closed geodesic in $M^3$.
  (A truncated cusp or leg that goes to $\infty$ is indicated
   by `$\rightarrow$'.)
 } % end-setcaption
 \centerline{\psfig{figure=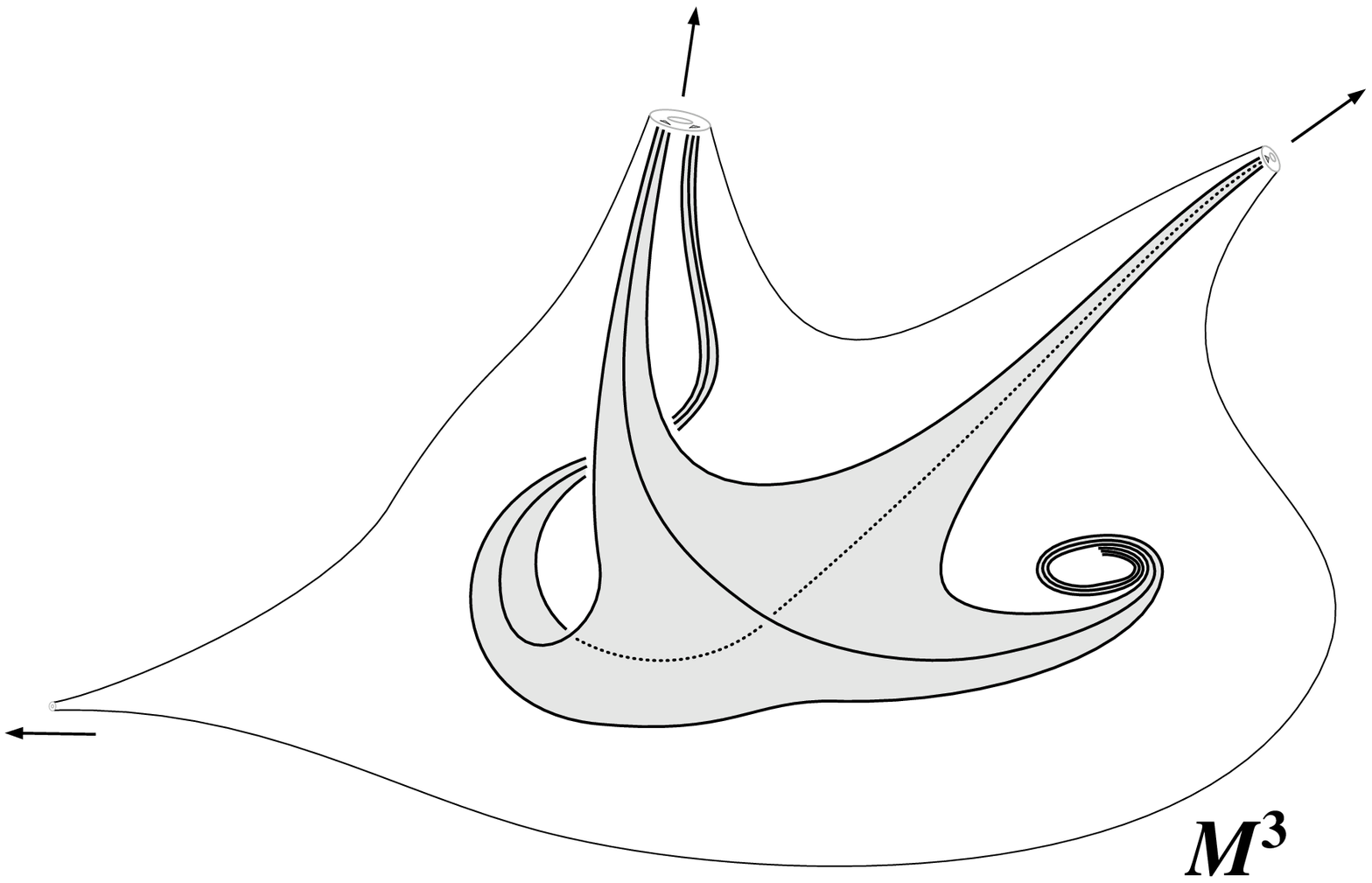,width=13cm,caption=}}
\end{figure}

\bigskip

\noindent $\bullet$
{\bf Dehn filling.} ([Ro] and [Th1].)
Given a $3$-manifold $N^3$ with a boundary component $\partial_0N^3$
a $2$-torus ${\Bbb T}^2$. One can fill $\partial_0N^3$ by sewing a
solid torus $D^2\times S^1$ to $N^3$ via a homeomorphism $h$ from
$\partial(D^2\times S^1)$ to $\partial_0N^3$. This procedure is
called a {\it Dehn-filling} of $N^3$. Define a {\it meridian} of
the solid torus to be any simple loop $\hat{m}$ in
$\partial(D^2\times S^1)$ that bounds in $D^2\times S^1$. Then the
topology of the new manifold $N^3\cup_h(D^2\times S^1)$ depends
only on the free homotopy class of the image $h(\hat{m})$ in
$\partial_0N^3$. For this reason, suppose that
$\partial N^3=\cup_{i=1}^k{\Bbb T}^2_i$ with fixed basis $(m_i,l_i)$
for $H_1({\Bbb T}^2_i,{\Bbb Z})$, then a manifold obtained from
$N^3$ by Dehn-filling can be denoted unambiguously by
$N^3_{(a_i,b_i;\,\cdots\,a_k,b_k)}$, where $(a_i,b_i)=\infty$ if
${\Bbb T}^2_i$ is not filled, or the relatively prime integer pair
from the equality $h_i(\hat{m})=a_im_i+b_il_i$ in
$H_1({\Bbb T}^2_i,{\Bbb Z})$ if ${\Bbb T}^2_i$ is filled.

\bigskip

\noindent $\bullet$
{\bf Knots and links in $S^3$.} ([Ki] and [Ro].)
Up to free homotopies in $S^3-K$, the {\it meridian} $m$ of a knot
$K$ in $S^3$ is the circle that bounds the fiber $2$-disk of the
normal bundle $\nu(K)$ of $K$ in $S^3$ and the {\it longitude}
$l$ of $K$ is a knot in $S^3$ that is parallel to $K$ and generates
the $0$-framing along $K$, as indicated in {\sc Figure 1-4}.
Note that the linking number $\lk(l,K)=0$, while $\lk(m,K)=1$ when
the orientations of $m$, $K$, and $S^3$ are chosen appropriately.
All the terminologies and facts used can be found in [Ro].
\begin{figure}[htbp]
\setcaption{{\sc Figure 1-4.}
\baselineskip 14pt
 The meridian $m$ (thick line) and the longitude $l$ (thin line)
 of a knot $K$ in $S^3$ (cf.\ {\sc Fig. 2.2} in [Ki]).
} % end-setcaption
\centerline{\psfig{figure=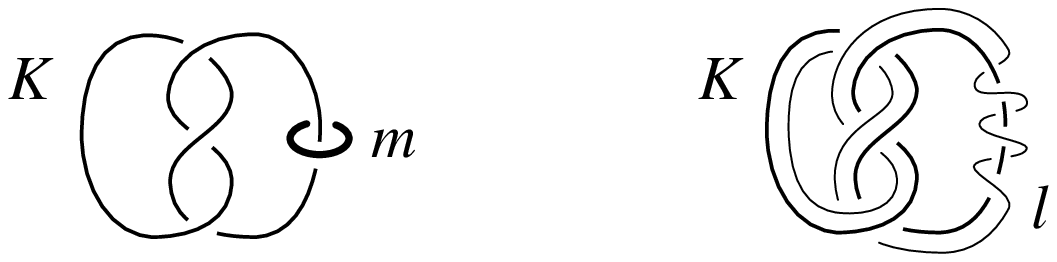,width=13cm,caption=}}
\end{figure}

\bigskip

\section{The construction of a class of K3T $7$-manifolds.}

By a {\it K3T $7$-manifold}, we mean a $7$-manifold that is fibred
over a hyperbolic $3$-manifold with generic fiber a K3 surface.
In this section, we construct a special class of such $7$-manifolds
which are either closed or with boundary component the Calabi-Yau
threefolds ${\Bbb T}^2\times\mbox{\rm K3}$.
Some necessary facts for the construction are summarized in
Sec.\ 2.1; the construction is given in Sec.\ 2.2; and examples are
provided in Sec.\ 2.3.

\bigskip

\subsection{The two ingredients:
   K3 surfaces with involution and hyperbolic 3-manifolds
   of finite volume.}

Some necessary facts for the construction are summarized in this
subsection. Readers are referred to [Bo], [G-W], [Ni4], [Vo] for
K3 surfaces with antisymplectic involution and to [B-P], [Th1] for
hyperbolic $3$-manifolds of finite volume.

\bigskip

\begin{flushleft}
{\bf K3 surfaces with antisymplectic involution..}
\end{flushleft}
Let $X$ be a K3 surface with an involution $\iota$ that acts by
$(-1)$ on the holomorphic $2$-form of $X$. Let $\itSigma$ be the set
of fixed points of $\iota$. Then $\itSigma$ is a disjoint union of
smooth complex curves in $X$ that fall into one of the following
three situations classified by Nikulin ([Ni4]):
\begin{quote}
 \hspace{-1.9em}(1)\hspace{1ex}
 $\itSigma$ is empty;

 \hspace{-1.9em}(2)\hspace{1ex}
  $\itSigma=C_1\cup C_1^{\prime}$, where $C_1$ and $C_1^{\prime}$
  are both elliptic curves;

 \hspace{-1.9em}(3)\hspace{1ex}
  $\itSigma=C_g+E_1+\,\cdots\,+E_k$. where $C_g$ is a curve of
  genus $g$ and $E_i$ are rational curves.
\end{quote}
And these complex curves descend isomorphically to complex curves
(also denoted by $\itSigma$) in
$X/\hspace{-.1ex}\mbox{\raisebox{-.4ex}{$\iota$}}$. In all cases,
the quotient
$X/\hspace{-.1ex}\mbox{\raisebox{-.4ex}{$\iota$}}$ is a smooth
complex surface.
The deformation space of the complex structures of such K3 surface
was discussed in [Dol] and [G-W] and will be addressed in Sec.\ 3.2.

\bigskip

\begin{flushleft}
{\bf A quick tour on hyperbolic 3-manifolds of finite volume.}
\end{flushleft}
The Mostow's rigidity theorem says that a complete hyperbolic
$3$-manifold $M^3$ of finite volume with a fixed fundamental group
is unique up to isometries. However, it turns out to be more
fruitful to think of $M^3$ as obtained from the canonical filling of
a compact hyperbolic $3$-manifold $N^3$ with toroidal boundary.
For example, one may take $N^3$ to be the thick part
$$
 M^3_{[\varepsilon,\infty)}\;
  =\;\{\,p\in M^3\,|\,
    \mbox{there is a ball of radius $\varepsilon$ embedded in $M^3$
     with center at $p$}\,\}
$$
of $M^3$ for some $\varepsilon$ small enough. While the hyperbolic
structure of $M^3$ does not allow deformation, the hyperbolic
structure on $N^3$ does and the deformation space has complex
dimension the number of components of $\partial N^3$ (if one neglects
the boundary behaviors). Furthermore, when $N^3$ is deformed
to another hyperbolic $3$-manifold ${N^3}^{\prime}$ with boundary,
${N^3}^{\prime}$ can be canonically filled and may lead to another
complete hyperbolic $3$-manifold ${M^3}^{\prime}$ of finite volume
with different topology. This gives a certain hierarchical structure
on the space ${\cal H}$ of complete hyperbolic $3$-manifolds of
finite volume. With an appropriate topology defined on ${\cal H}$,
the volume function $\vol$ on ${\cal H}$ is continuous and
finite-to-one. Its image is a countable well-ordered subset
(i.e.\ every subset has a smallest element) in ${\Bbb R}_{>0}$.
These are results in the so-called
{\it J$\phi$rgensen-Thurston theory} of hyperbolic $3$-manifold of
finite volume. For our purpose, let us explain more detail of the
canonical filling. The deformation space of hyperbolic structures on
$N^3$ will be addressed in Sec.\ 3.3.

Recall the upper half-space model ${\Bbb C}\times{\Bbb R}_+$ for
${\Bbb H}^3$ from Sec.\ 1. Let $N^3$ be a compact hyperbolic
$3$-manifold with toroidal boundary. The hyperbolic structure on
$N^3$ induces a holonomy map
$$
 \mu\;:\;\pi_1(N^3)\;\longrightarrow\;\PSL(2;{\Bbb C})\,.
$$
Let ${\Bbb T}^2$ be a boundary component of $N^3$. Then
$\mu(\pi_1({\Bbb T}))$ is an abelian subgroup in $\PSL(2;{\Bbb C})$.
Thus elements in $\mu(\pi_1({\Bbb T}^2))$ must share the same set
of fixed points on $S^2_{\infty}$. Furthermore the number of fixed
points can only be either one or two.

If $\mu(\pi_1({\Bbb T}^2))$ has only one fixed point on
$S^2_{\infty}$, after conjugation one may assume that it is $\infty$.
Then $\mu(\pi_1({\Bbb T}^2))$ is a rank $2$ lattice in the
translation group of ${\Bbb C}$. The corresponding boundary of $N^3$
can be filled by the infinite cusp ${\Bbb T}^2\times{\Bbb R}_+$,
geometrically modelled on the quotient
$$
 \{(z,t)\,|\,z\in{\Bbb C}, t\ge h\}/
 \hspace{-.1ex}\mbox{\raisebox{-.4ex}{$z\sim z+c_1, z\sim z+c_2$}}\,
$$
where $\{z\sim z+c_1, z\sim z+c_2\}$ is a set of generators
of $\mu(\pi_1({\Bbb T}^2))$ and $h$ depends by the hyperbolic
structure of $N^3$ around that boundary ({\sc Figure 2-2}(a)).
Note that this gives an Euclidean structure on the corresponding
${\Bbb T}^2$-boundary of $N^3$.

If $\mu(\pi_1({\Bbb T}^2))$ has two fixed points on $S^2_{\infty}$,
after conjugation one may assume that they are $0$ and $\infty$.
Then $\mu(\pi_1({\Bbb T}^2))$ is an abelian subgroup in
$\GL(1;{\Bbb C})={\Bbb C}^{\times}$. The lifted holonomy map of
$\mu|_{\pi_1({\scriptsizeBbb T}^2)}$
$$
 \widetilde{\mu}\;:\; \pi_1({\Bbb T}^2)\;\longrightarrow\;
    \widetilde{{\Bbb C}^{\times}}\,,
$$
where $\widetilde{{\Bbb C}^{\times}}$ is the universal covering
group of ${\Bbb C}^{\times}$, is now discrete and injective and
gives an complex affine structure on the corresponding
${\Bbb T}^2$-boundary of $N^3$. Let ${\cal N}_{\varepsilon}$ be the
closed $\varepsilon$-neighborhood of the $t$-axis in ${\Bbb H}^3$
with the $t$-axis deleted and $\widetilde{{\cal N}_{\varepsilon}}$
be the universal covering of ${\cal N}_{\varepsilon}$, equipped with
the lifted hyperbolic structure. Then
$\widetilde{\mu}(\pi_1({\Bbb T}^2))$ acts freely on
$\widetilde{{\cal N}_{\varepsilon}}$ as a group of isometries.
This action extends to the metric completion
$\overline{\widetilde{{\cal N}_{\varepsilon}}}$ of
$\widetilde{{\cal N}_{\varepsilon}}$. The corresponding boundary
of $N^3$ can then be filled by 
$\overline{\widetilde{{\cal N}_{\varepsilon}}}/
  \hspace{-.1ex}\mbox{\raisebox{-.4ex}{$\widetilde{\mu}(
                                         \pi_1({\Bbb T}^2))$}}$
for some $\varepsilon$ depending on the hyperbolic structure of
$N^3$ around that boundary. ({\sc Figure 2-1}.)
\begin{figure}[htbp]
 \setcaption{{\sc Figure 2-1.}
 \baselineskip 14 pt
  The deleted neifhboorhood ${\cal N}_{\varepsilon}$ of $t$-axis
  and its universal covering $\widetilde{{\cal N}_{\varepsilon}}$,
  (for clarity, only their boundary is shown).
  A fundamental domain of $\pi_1({\Bbb T}^2)$-action on
  $\partial\widetilde{{\cal N}_{\varepsilon}}$ via $\widetilde{\mu}$
  and its image on $\partial{\cal N}_{\varepsilon}$ are indicated.
 } % end-setcaption
 \centerline{\psfig{figure=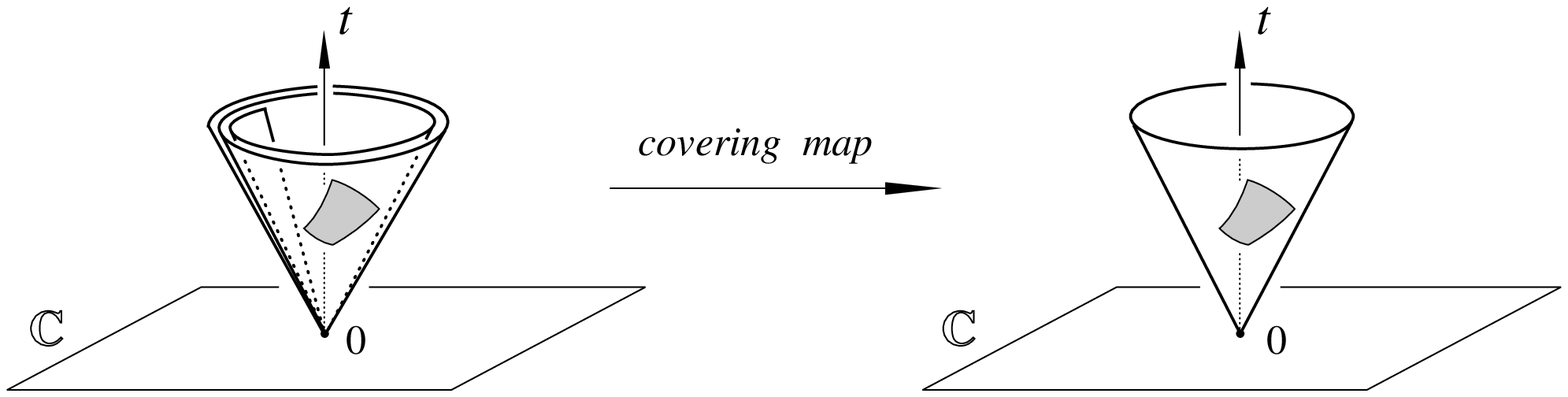,width=13cm,caption=}}
\end{figure}

After performing the above canonical filling to every boundary of
$N^3$, the resulting $3$-space ${W^3}$ is hyperbolic of finite
volume and metrically complete. However, it may acquire some
non-manifold point or curvature defect from the second type of
filling. Let us next turn to these possible singularities.

The completion $\overline{\widetilde{{\cal N}_{\varepsilon}}}$
of $\widetilde{{\cal N}_{\varepsilon}}$ is obtained by adding 
the (lifted) $t$-axis. The induced action of
$\widetilde{\pi_1({\Bbb T}^2)}$ on the $t$-axis induces a group
homomorphism
$$
 \hat{\mu}\;:\;\pi_1({\Bbb T}^2)\; \longrightarrow\;
                                     ({\Bbb R}_{>0}, \times)\,.
$$
There are two situations:
\begin{itemize}
 \item                    
 {\it Case (i)$\,$}:
 $\hat{\mu}(\pi_1({\Bbb T}^2))$ {\it is dense in} ${\Bbb R}_{>0}$.
 Then
 $\overline{\widetilde{{\cal N}_{\varepsilon}}}/
   \hspace{-.1ex}\mbox{\raisebox{-.4ex}{$\widetilde{\mu}(
                                          \pi_1({\Bbb T}^2))$}}$
 is the one-point compactification of
 $\widetilde{{\cal N}_{\varepsilon}}/
   \hspace{-.1ex}\mbox{\raisebox{-.4ex}{$\widetilde{\mu}(
                                          \pi_1({\Bbb T}^2))$}}$.
 The corresponding boundary in $N^3$ is filled by a cone over
 ${\Bbb T}^2$ ({\sc Figure 2-2}(b)).

 \item
 {\it Case (ii)$\,$}:
 $\hat{\mu}(\pi_1({\Bbb T}^2))$ {\it is discrete in} ${\Bbb R}_{>0}$.
 Then the completion
 $\overline{\widetilde{{\cal N}_{\varepsilon}}}/
   \hspace{-.1ex}\mbox{\raisebox{-.4ex}{$\widetilde{\mu}(
                                          \pi_1({\Bbb T}^2))$}}$
 is topologically a solid torus. The difference
 $\overline{\widetilde{{\cal N}_{\varepsilon}}}/
   \hspace{-.1ex}\mbox{\raisebox{-.4ex}{$\widetilde{\mu}(
                                          \pi_1({\Bbb T}^2))$}}
 -\widetilde{{\cal N}_{\varepsilon}}/
   \hspace{-.1ex}\mbox{\raisebox{-.4ex}{$\widetilde{\mu}(
                                          \pi_1({\Bbb T}^2))$}}$
 is the loop
 $\gamma=\mbox{$t$-axis}\,/\hspace{-.1ex}\mbox{\raisebox{-.4ex}{$
                                   \hat{\mu}(\pi_1({\Bbb T}^2))$}}$.
 Let $\lambda_1$ generate the kernel of $\hat{\mu}$ and $\lambda_2$
 generate the image of $\hat{\mu}$. Then $\gamma$ has length
 $|\log|\lambda_2||$ and the normal cross section of $\gamma$ in
 the completion is a $2$-dimensional hyperbolic cone $D^2_{\theta}$
 of cone angle $\theta=|\arg\lambda_1|$
 ({\sc Figure 2-2}(c) and (d)). Hence, when $\theta\ne 2\pi$,
 there are curvature defects along $\gamma$.
 \end{itemize}
\begin{figure}[htbp]
\setcaption{{\sc Figure 2-2.}
\baselineskip 14pt
 The boundary of a compact hyperbolic $3$-manifold $N^3$ with
 toroidal boundary can be canonically filled by 
 (a) a cusp, (b) a cone,
 (c) a solid torus with curvature defects along the core, or
 (d) a nice hyperbolic solid torus. The result is denoted by $W^3$
 in the figure.
 In Cases (c) and (d), the core loop is indicated by a grey loop.  
} % end-setcaption
\centerline{\psfig{figure=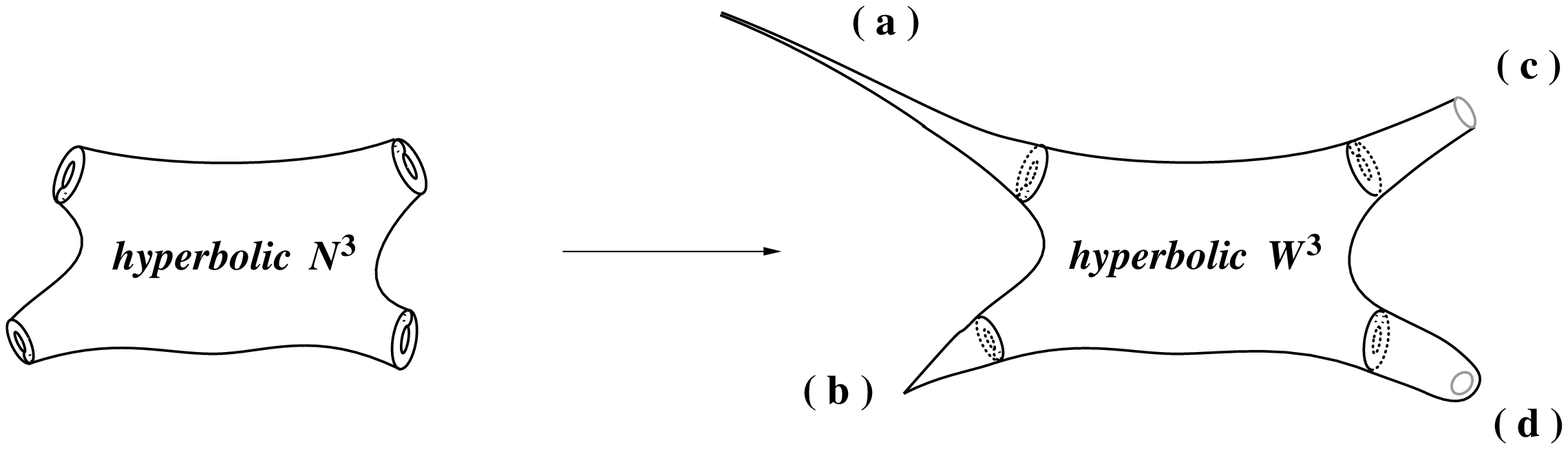,width=13cm,caption=}}
\end{figure}

\bigskip

\noindent
{\it Remark 2.1.1 [hyperbolic Dehn surgery]}.
Recall the procedure
$$
 M^3\;\rightarrow\; N^3\; \rightarrow\;\mbox{deforming $N^3$}\;
 \rightarrow\;\mbox{canonical filling}\; \rightarrow\;{M^3}^{\prime}\,.
$$
from a complete hyperbolic $M^3$ of finite volume
at the start of the tour.
Assume that the deformed $N^3$ is filled either by infinite cusps or
solid torus in Case (ii) with $\theta=2\pi$. Then the underlying
topology of ${M^3}^{\prime}$ is a $3$-manifold obtained from $N^3$ by
performing a Dehn filling on the boundary. Such ${M^3}^{\prime}$ is
said to be obtained from $M^3$ by a hyperbolic surgery.
The following fact is due to Thurston ([Th1], see also [B-P]):

\bigskip

\noindent
{\bf Fact 2.1.1.1 [hyperbolic surgery theorem].} {\it
 Let $M^3$ be a complete hyperbolic $3$-manifold of finite volume
 with $h$ cusps. Fixed a basis $(m_i,l_i)$ for
 $H_1({\Bbb T}^2_i; {\Bbb Z})$, where ${\Bbb T}^2_i$ is the
 $2$-torus associated to the $i$-th cusp.
 Let $M^3_{(p_1,q_1;\,\cdots\,; p_h,q_h)}$ be the $3$-manifold
 obtained by performong a $(p_i,q_i)$-Dehn surgery on the $i$-th
 cusp, where $(p_i,q_i)$ is a pair of coprime integers or the
 symbol $\infty$ if the $i$-th cusp is left unsurgered. Then
 $M^3_{(p_1,q_1;\,\cdots\,; p_h,q_h)}$ admits a hyperbolic
 structure if all $(p_i, q_i)$ are close to $\infty$ in
 $({\Bbb Z}\oplus{\Bbb Z})\cup\{\infty\}$.
} % end-fact

\bigskip

\noindent
This is an important theorem that gives a hierarchical structure
on th space ${\cal H}$ of hyperbolic $3$-manifolds of finite volume
and is good to keep in mind when considering such $3$-manifolds.

\bigskip

\subsection{The construction of K3T 7-manifolds by bundle-filling.}

Let $M^3$ be a complete hyperbolic $3$-manifold of finite volume,
$X$ be a K3 surface with antisymplectic involution $\iota$, and
$\itSigma$ be the set of fixed points of $\iota$. Let
$\itGamma=\pi_1(M^3)$ and
$\rho_X:\itGamma\rightarrow\langle\,\iota\,\rangle={\Bbb Z}_2$
be a representation. Then the universal covering $\widetilde{M^3}$
of $M^3$ is ${\Bbb H}^3$ and $(\itGamma, \rho_X(\Gamma))$ acts
freely on ${\Bbb H}^3\times X$ by the diagonal action. Consequently,
one obtains a K3T $7$-manifold
$$
 \pi\;:\;
 M^7\, =\, M^3\times_{\rho_X}X\,
  =\,({\Bbb H}^3\times X)/
     \hspace{-.2ex}\mbox{\raisebox{-.4ex}{$(\itGamma,
                             \rho_X(\itGamma))$}}\;
                         \longrightarrow\; M^3\,.
$$
In this case, $\pi$ is a K3-bundle over $M^3$. Note that $\rho_X$
can be regarded as an element in $H^1(M^3;{\Bbb Z}_2)$ and vice versa.
The isometry group of $M^3$ (in general it is trivial) acts on
$H^1(M^3;{\Bbb Z}_2)$, the quotient labels then a class of K3T
$7$-manifolds with fiber $X$ that are naturally associated to $M^3$.
Thurston's work on hyperbolic $3$-manifolds of finite volume suggests
a way to construct a family of K3T descendants from any such $\pi$
by ``bundle-filling", as we shall now explain.

\bigskip

\begin{flushleft}
{\bf The basic idea.}
\end{flushleft}
Regard $M^3$ as a filling of some compact $N^3$ with toroidal
boundary. The hyperbolic structure on $N^3$ can be regarded as a
collection of local hyperbolic charts
$$
 \{\,\varphi_{\alpha}\,:\,U_{\alpha}\,\longrightarrow\,{\Bbb H}^3\,
  |\, \alpha\in\,\mbox{some index set $I$}\,\}\,,
$$
where $\{U_{\alpha}\,|\,\alpha\in I\}$ is a covering of $N^3$.
Deformation of hyperbolic structures on $N^3$ can then be thought of
as shufflings of these local charts. This leads to alteration of the
transition functions $\{\varphi_{\alpha\beta}\}$ between charts and
also the corresponding representation
$\pi_1(N^3)=\itGamma\rightarrow \PSL(2;{\Bbb C})$.

Now choose $\{U_{\alpha}\,|\,\alpha\in I\}$ to be also a local
trivialization of $\pi$. Let
$(\varphi_{\alpha\beta},h_{\alpha\beta})$ be the transition function
between $U_{\alpha}\times X$ and $U_{\beta}\times X$. Then, by
construction, one can choose $h_{\alpha\beta}$ to be either the
identity map $\Id_X$ or the involution $\iota$ on $X$.
When one deforms the hyperbolic structure on $N^3$,
$\varphi_{\alpha\beta}$ get changed but the corresponding
$h_{\alpha\beta}$ can be left the same. This leads to a new
bundle $\pi^{\prime}$ topologically equivalent to $\pi|_{N^3}$.
Let ${M^3}^{\prime}$ be the filling of the deformed $N^3$, following
Thurston. As will be explained below, the filling of $N^3$ to
${M^3}^{\prime}$ induces a filling of $\pi|_{N^3}$ to a K3-fibration
$\pi^{\prime}:W^7\rightarrow{M^3}^{\prime}$ in a way that depends
only on the hyperbolic structure on $N^3$ and $\rho_X$.
By understanding these fillings and how the resulting singularities
of $W^7$ can be resolved, one obtains a K3T $7$-manifold
${M^7}^{\prime}$ that descends from $\pi$.

\bigskip

\begin{flushleft}
{\bf The fibration over a boundary ${\Bbb T}^2$ of $N^3$ and its
     filling.}
\end{flushleft}
With notations from earlier discussions, let ${\Bbb T}^2$ be a
boundary component of the deformed $N^3$.
Recall $\mu$, $\widetilde{\mu}$, and $\hat{\mu}$ from Sec.\ 2.1.

If $\mu(\pi_1({\Bbb T}^2))$ has only one fixed point on
$S^2_{\infty}$, this boundary ${\Bbb T}^2$ is filled to a cusp
${\cal C}^{\prime}={\Bbb T}^2\times{\Bbb R}_+$ in ${M^3}^{\prime}$.
Let ${\cal C}$ be the corresponding cusp in $M^3$, then the
restriction of $\pi^{\prime}$ to a collar of ${\Bbb T}^2$ in $N^3$
extends to $\pi^{\prime}|_{{\cal C}^{\prime}}$ that is the pullback
of $\pi|_{\cal C}$ via a quasi-isometry from ${\cal C}^{\prime}$ to
${\cal C}$. In this case, $\pi^{\prime}|_{{\cal C}^{\prime}}$ is
topologically equivalent to $\pi|_{\cal C}$.

If $\mu(\pi_1({\Bbb T}^2))$ has two fixed points on $S^2_{\infty}$,
then the restriction of $\pi^{\prime}$ over a collar of ${\Bbb T}^2$
in $N^3$ extends to
$$
 \pi^{\prime}|_{
      \overline{\widetilde{{\cal N}_{\varepsilon}}}/
       \hspace{-.1ex}\mbox{\scriptsize
        \raisebox{-.4ex}{$\widetilde{\mu}(
                \pi_1({\scriptsizeBbb T}^2))$}}     }\;:\;
   \{\overline{\widetilde{{\cal N}_{\varepsilon}}}
    \times X\}/ \hspace{-.1ex}\mbox{\raisebox{-.4ex}{$(
     \widetilde{\mu}(\pi_1({\Bbb T}^2)),\,
                      \rho_X(\pi_1({\Bbb T}^2)))$}}\;
   \longrightarrow\;
   \overline{\widetilde{{\cal N}_{\varepsilon}}}/
     \hspace{-.1ex}\mbox{\raisebox{-.4ex}{$\widetilde{\mu}(
                                        \pi_1({\Bbb T}^2))$}}\,,
$$
where the group action is diagonal. Corresponding to Cases (i) and
(ii) in Sec.\ 2.1, one has respectively
\begin{itemize}
 \item
 {\it Case (i$\,^{\prime}$)$\,$}:
 $\hat{\mu}(\pi_1({\Bbb T}^2))$ {\it is dense in} ${\Bbb R}_{>0}$.
 Recall that 
 $\overline{\widetilde{{\cal N}_{\varepsilon}}}/
    \hspace{-.1ex}\mbox{\raisebox{-.4ex}{$\widetilde{\mu}(
                                           \pi_1({\Bbb T}^2))$}}$
 is the one-point compactification of
 $\widetilde{{\cal N}_{\varepsilon}}/
   \hspace{-.1ex}\mbox{\raisebox{-.4ex}{$\widetilde{\mu}(
                                           \pi_1({\Bbb T}^2))$}}$
 by, say, $p_{\ast}$. Then the fiber of $\pi^{\prime}$ over
 $p_{\ast}$ is
 $X_{\ast}=X/\hspace{-.1ex}\mbox{\raisebox{-.4ex}{$\rho_X(
                                            \pi_1({\Bbb T}^2))$}}$.
 If $\rho_X(\pi_1({\Bbb T}^2))=\langle\iota\rangle$,
 then $X_{\ast}=X/\hspace{-.1ex}\mbox{\raisebox{-.4ex}{$\iota$}}$
 is an exceptional fiber of multiplicity $2$.
 If $\rho_X(\pi_1({\Bbb T}^2))=\{Id_X\}$, then the extended
 $\pi^{\prime}$ over
 $\overline{\widetilde{{\cal N}_{\varepsilon}}}/
   \hspace{-.1ex}\mbox{\raisebox{-.4ex}{$\widetilde{\mu}(
                                          \pi_1({\Bbb T}^2))$}}$     
 is a trivial fibration and $X_{\ast}=X$ is a regular fiber.
 
 \item
 {\it Case (ii$\,^{\prime}$)$\,$}:
 $\hat{\mu}(\pi_1({\Bbb T}^2))$ {\it is discrete in} ${\Bbb R}_{>0}$.
 Recall the core loop $\gamma$ of
 $\overline{\widetilde{{\cal N}_{\varepsilon}}}/
   \hspace{-.1ex}\mbox{\raisebox{-.4ex}{$\widetilde{\mu}(
                                          \pi_1({\Bbb T}^2))$}}$.
 Then the fiber of $\pi^{\prime}$ over $\gamma$ is
 $X_c=X/\hspace{-.1ex}\mbox{\raisebox{-.4ex}{$\rho_X(\lambda_1)$}}$.
 If $\rho_X(\lambda_1)=\iota$, then
 $X_c=X/\hspace{-.1ex}\mbox{\raisebox{-.4ex}{$\iota$}}$ is an
 exceptional fiber of multiplicity $2$.
 If $\rho_X(\lambda_1)=\Id$, then $X_c=X$ is a regular fiber.
\end{itemize}

Thus, after filling $\pi^{\prime}|_{N^3}$ over all the boundary
${\Bbb T}^2$ of $N^3$, one obtains a flat K3-fibration
$$
 \pi^{\prime}\::\:W^7\: \longrightarrow\: {M^3}^{\prime}\,.
$$
By construction, if one regards $M^3$ as embedded in
${M^3}^{\prime}$, then the monodromy of $\pi^{\prime}|_{M^3}$ is
the same as the monodromy of $\pi$. This $W^7$ in general is not
a manifold. Let us now turn to its singularities.

\bigskip

\begin{flushleft}
{\bf The singular locus of $W^7$ and its resolution.}
\end{flushleft}
Let $S$ be the singular set $S$ of $W^7$, that consists of all the
non-manifold points of $W^7$. Then, from previous discussion, $S$
sits only over the difference ${M^3}^{\prime}-M^3$, which is a
disjoint union of finitely many points $p_{\ast}$ and simple loops
$\gamma$ obtained from filling the deformed hyperbolic structure of
$N^3$. Let $S_0$ be a component of $S$ and $\nu(\,\cdot\,)$ be the
tubular neighborhood of a subset in $W^7$. We shall discuss first
the topology of $\nu({\pi^{\prime}}^{-1}(p_{\ast}))$ and
$\nu({\pi^{\prime}}^{-1}(\gamma))$, and then $S_0$ and how $S_0$
may be resolved.

In the following discussions, ${\Bbb T}^2$ is the boundary component 
of $N^3$ involved. Also, we shall denote
{\it the cone over a base $B$} by $\Cone(B)$ and the
{\it $2$-dimensional hyperbolic cone with cone angle $\theta$}
by $D^2_{\theta}$.

The two cases are as follows:

\medskip

\noindent \hspace{.5ex}
{\it Case (a)}$\,$:
{\it Over $p_{\ast}$ in Case (i$\,^{\prime}$) above.}
There are two situations:
\begin{enumerate}
 \item
 $\rho_X(\pi_1({\Bbb T}^2))=\{\Id_X\}$.
 Then ${\pi^{\prime}}^{-1}(p_{\ast})=X$ and
 $\nu({\pi^{\prime}}^{-1}(p_{\ast}))=X\times \Cone({\Bbb T}^2)$.
 Thus $S_0={\pi^{\prime}}^{-1}(p_{\ast})=X$.

 \item
 $\rho_X(\pi_1({\Bbb T}^2))=\langle\iota\rangle$.
 Then
 ${\pi^{\prime}}^{-1}(p_{\ast})
        =X/\hspace{-.1ex}\mbox{\raisebox{-.4ex}{$\iota$}}$.
 Let $\lambda_0\in\pi_1({\Bbb T}^2)$ generate the
 image of $\rho_X$. Then there is a double covering
 $\kappa:\hat{\Bbb T}^2\rightarrow {\Bbb T}^2$ such that
 $\rho_X\circ\kappa_{\ast}(\pi_1(\hat{\Bbb T}^2))$ is trivial.
 Let $h$ be the nontrivial deck transformation on $\hat{\Bbb T}^2$
 associated to $\kappa$. Then
 $\partial\nu({\pi^{\prime}}^{-1}(p_{\ast}))=
  (\hat{\Bbb T}^2\times X)/
   \hspace{-.1ex}\mbox{\raisebox{-.4ex}{$(h,\iota)$}}$,
 which is fibered over ${\pi^{\prime}}^{-1}(p_{\ast})$ with generic
 fiber $\hat{\Bbb T}^2$ and exceptional fiber
 ${\Bbb T}^2=\hat{\Bbb T}^2/
                   \hspace{-.1ex}\mbox{\raisebox{-.4ex}{$h$}}$
 of multiplicity $2$ over $\itSigma$. Consequently, 
 $\nu({\pi^{\prime}}^{-1}(p_{\ast}))$ is a
 $\Cone(\hat{\Bbb T}^2)$-fibration over
 $X/\hspace{-.1ex}\mbox{\raisebox{-.4ex}{$\iota$}}$.
 with exceptional $\Cone({\Bbb T}^2)$-fiber of multiplicity $2$
 over $\itSigma$. The monodromy of the fibration around the loop
 $[\iota]$, which generates the orbifold fundamental group
 $\pi_1^{\rm orb}(S_0)$, is the extension of $h$ on $\hat{\Bbb T}^2$
 to $\Cone(\hat{\Bbb T}^2)$. Thus
 $S_0={\pi^{\prime}}^{-1}(p_{\ast})
     =X/\hspace{-.1ex}\mbox{\raisebox{-.4ex}{$\iota$}}$.
\end{enumerate}

\bigskip

\noindent\hspace{.5ex}
{\it Case (b)}$\,$:
{\it Over $\gamma$ in Case (ii$\,^{\prime}$) above.}
There are three situations:
\begin{enumerate}
 \item
 $\rho_X(\pi_1({\Bbb T}^2))=\{\Id_X\}\,$:
 Then ${\pi^{\prime}}^{-1}(\gamma)=X\times\gamma$ and
 $\nu({\pi^{\prime}}^{-1}(\gamma))
   =(X\times\gamma)\times D^2_{|\arg\lambda_1|}$.
 Thus $S_0$ is empty.

 \item
 $\rho_X(\lambda_1)=\iota$ and $\rho_X(\lambda_2)=\Id_X\,$: 
 Then
 ${\pi^{\prime}}^{-1}(\gamma)
   =X/\hspace{-.1ex}\mbox{\raisebox{-.4ex}{$\iota$}}\times\gamma$.
 Let
 ${\cal C}=\overline{\widetilde{{\cal N}_{\varepsilon}}}/
      \hspace{-.1ex}\mbox{\raisebox{-.4ex}{$\widetilde{\mu}(
                                         \pi_1({\Bbb T}^2))$}}$.
 Then there is a double branched covering
 $\kappa:\hat{\cal C}\rightarrow{\cal C}$
 branched over the core $\gamma$ of ${\cal C}$ such that the
 pullback fibration $\kappa^{\ast}(\pi^{\prime}|_{\cal C})$
 is trivial over $\hat{\cal C}$. The total space of
 $\kappa^{\ast}(\pi^{\prime}|_{\cal C})$ is $\hat{\cal C}\times X$.
 Let $h$ be the nontrivial deck transformation on $\hat{\cal C}$
 associated to $\kappa$. Since $\nu({\pi^{\prime}}^{-1}(\gamma))$
 is the total space of $\pi^{\prime}|_{\cal C}$, one has
 $\nu({\pi^{\prime}}^{-1}(\gamma))=(\hat{\cal C}\times X)/
               \hspace{-.1ex}\mbox{\raisebox{-.4ex}{$(h,\iota)$}}$,
 where the action is diagonal. Thus
 $\nu({\pi^{\prime}}^{-1}(\gamma))$ fibers
 over ${\pi^{\prime}}^{-1}(\gamma)$ with generic fiber
 $D^2_{2|\arg\lambda_1|}$. The exceptional
 fibers lie over $\itSigma\times\gamma$ and each is isometric to
 $D^2_{|\arg\lambda_1|}$, with multiplicity $2$. This shows
 that $\nu({\pi^{\prime}}^{-1}(\gamma))$ has transverse
 $A_1$-singularities along $\itSigma\times\gamma$ and that 
 $S_0=\itSigma\times\gamma$.

 \item
 $\rho_X(\lambda_1)=\Id_X$ and $\rho_X(\lambda_2)=\iota\,$:
 Then ${\pi^{\prime}}^{-1}(\gamma)=X_{\iota}$, the mapping torus
 of $X$ associated to $\iota$. Analogous to the previous situation,
 there is a double covering
 $\kappa:\hat{\cal C}\rightarrow{\cal C}$,
 induced by a double covering $S^1\times\gamma$,
 such that the pullback fibration
 $\kappa^{\ast}(\pi^{\prime}|_{\cal C})$ is trivial over
 $\hat{\cal C}$. The total space of
 $\kappa^{\ast}(\pi^{\prime}|_{\cal C})$
 is $\hat{\cal C}\times X$. Let $h$ be the nontrivial
 deck transformation on $\hat{\cal C}$ associated to
 $\kappa$. Then
 $\nu({\pi^{\prime}}^{-1}(\gamma))=(\hat{\cal C}\times X)/
            \hspace{-.1ex}\mbox{\raisebox{-.4ex}{$(h,\iota)$}}$,
 where the action is diagonal. Thus
 $\nu({\pi^{\prime}}^{-1}(\gamma))$ is a regular fiberation over
 ${\pi^{\prime}}^{-1}(\gamma)$ with fiber $D^2_{|\arg\lambda_1|}$
 and $S_0$ is empty.  \\[-1ex]

 Note that, for the situation 
 $\rho_X(\lambda_1)=\rho_X(\lambda_2)=\iota$, one can replace
 $\lambda_2$ by $\lambda_2-\lambda_1$ and render it Situation (2)
 above. 
\end{enumerate}

\bigskip

Thus, if Case (i$^{\prime}$) happens for some boundary of $N^3$,
then, from Case (a), $W^7$ can never be a manifold. Nor is there
any known standard way to resolve such singularities.

On the other hand, if Case (ii$^{\prime}$) happens for all the
boundary components of $N^3$, both Case (b-1) and Case (b-3) above
lead only to manifold-points in $W^7$, while the $A_1$-singularities
in Case (b-2) can be resolved by transverse blowups along
$S_0=\itSigma\times\gamma$ (cf.\ Remark 2.2.1).
After, resolving the singularities, one obtains then a K3T
$7$-manifolds
$$
\widetilde{\pi}^{\prime}\;:\; {M^7}^{\prime} \;
                          \longrightarrow\; {M^3}^{\prime}\,.
$$
When Case (b-2) happens, the exceptional fiber of
$\widetilde{\pi}^{\prime}$ over a point in the corresponding
$\gamma$ is then
$X/\hspace{-.1ex}\mbox{\raisebox{-.4ex}{$\iota$}}\,
                                     \cup\, (\itSigma\times\CP^1)$,
where the two components intersect along $\itSigma$.
Over the complement of such $\gamma$, $\widetilde{\pi}^{\prime}$
is the same as $\pi^{\prime}$.

\bigskip

\noindent
{\it Remark 2.2.1 [hierarchy of K3T].}{
For a fixed $(X,\iota)$ and $\rho_X$, the space $W^7$ obtained by 
bundle-filling is determined by the hyperbolic structure on $N^3$.
Thus the associated space of K3T $7$-manifolds also exhibit a
hierarchical structure inherited from that on the space of hyperbolic
$3$-manifolds of finite volume. However, notice that, when Case (b-2)
happens, the set of isotopy classes of identifications of
$\hat{\cal C}$ with $\gamma\times{\Bbb C}$ is parametrized by
$\pi_1(\GL(1;{\Bbb C}))={\Bbb Z}$. Non-isotopic identifications may
lead to non-homeomorphic ${M^7}^{\prime}$ from the same $W^7$.
} % end_remark

\bigskip

\subsection{Examples from the link complements in $S^3$.}
In [Th4], Thurston showed that the interior of a compact $3$-manifold
$M^3$ is hyperbolic if and only if $M^3$ is prime, homotopically
atoroidal, and not homeomorphic to the quotient
$({\Bbb T}^2\times I)/
   \hspace{-.1ex}\mbox{\raisebox{-.4ex}{${\Bbb Z}_2$}}$,
where the ${\Bbb Z}_2$ acts on ${\Bbb T}^2$ as the covering group
of ${\Bbb T}^2$ over the Klein bottle and on the interval $I=[0,1]$
by the reflection with respect to $\frac{1}{2}$. A corollary of this
is that a knot $K$ in $S^3$ is hyperbolic if and only if $K$ is
neither a satellite nor a torus knot ([Th2]). Combined with
Fact 2.1.1.1 in Sec.\ 2.1 and the construction in Sec.\ 2.2,
this shows that there are abundant of nontrivial closed K3T
$7$-manifolds.

To construct nontrivial K3T $7$-manifolds that have the Calabi-Yau
threefold ${\Bbb T}^2\times X$ as a boundary component, let $L$ be
a link of $k$ many components in $S^3$ such that its complement
$S^3-L$ admits a complete hyperbolic structure. From the theorem of
Thurston mentioned in the beginning of the previous paragraph, there
are plenty of such $L$ in $S^3$ and their complement provide us with
basic examples of complete hyperbolic $3$-manifolds of finite volume.
It is a basic fact, following the Alexander duality and the
universal coefficient theorem [Mun], that
$H^1(S^3-L;{\Bbb Z}_2)=\oplus_k\,{\Bbb Z}_2$
is generated by taking the ${\Bbb Z}_2$-reduction of the linking
number with respect to a component of $L$. Thus there are $2^k$-many
homomorphisms $\rho_X$ from $\pi_1(S^3-L)$ to $\langle\iota\rangle$.
From the Mostow's rigidity theorem, unless $S^3-L$ admits a
nontrivial group of isometries, distinct $\rho_X$ gives rise to
non-isomorphic K3-bundles
$$
 \pi\;:\: (S^3-L)\times_{\rho_X} X\;\longrightarrow\; M^3=S^3-L\,.
$$
Applying the deformation and
filling in Sec.\ 2.2 to $\pi$, one generates then many other
examples of K3T $7$-manifolds. By choosing $\rho_X$ so that its
restriction to the unsurgered ${\Bbb T}^2$-boundary of $S^3-L$
is trivial, one then obtains many examples of nontrivial K3T
$7$-manifolds with boundary component ${\Bbb T}^2\times X$.

Let us now give some examples to illuminate the discussions.
The detail of the hyperbolic structure of the knot/link complements
that appear in these examples is in [Th1].

\bigskip

\noindent
{\bf Example 2.3.1 [closed K3T via figure-8 knot].}
Let $K$ be the {\it figure-8 knot} in $S^3$, as shown in
{\sc Figure 2-3-1}. 
\begin{figure}[htbp]
\setcaption{{\sc Figure 2-3-1.}
\baselineskip 14pt
 The figure-8 knot in $S^3$.
} % end-setcaption
\centerline{\psfig{figure=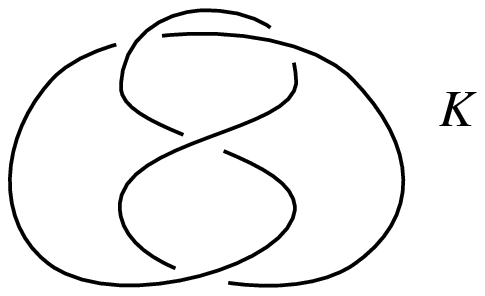,width=13cm,caption=}}
\end{figure}
\newline
Then $S^3-K$ is hyperbolic and $H^1(S^3-K;{\Bbb Z}_2)={\Bbb Z}_2$.
Let $m$ be the meridian and $l$ be the longitude of $K$.
Thurston proves that (Theorem 4.7 in [Th1]) every manifold obtained
by Dehn surgery along figure-8 knot has a hyperbolic structure,
except the six manifolds $(S^3-K)_{(\pm a,\pm b)}$ where
$(a,b)=(1,0)$, $(0,1)$, $(1,1)$, $(2,1)$, $(3,1)$, or $(4,1)$.

The two elements in $H^1(S^3-K;{\Bbb Z}_2)$ are
$$
\begin{array}{rcllr}
  \rho_0 & : & m\;\longmapsto\;\Id_X\,,
             & (\,l\;\longmapsto\;\Id_X\,) & \mbox{and}\\[1ex]
  \rho_1 & : & m\;\longmapsto\;\;\;\iota\,,
             & (\,l\;\longmapsto\;\Id_X\,)\,. &
\end{array}
$$
For $\rho_0$, $(S^3-K)\times_{\rho_0}X=(S^3-K)\times X$ and the
deform-and-fill procedure applied to $(S^3-K)\times X$ only yields
K3T $7$-manifolds in the product form ${M^3}^{\prime}\times X$.

For $\rho_1$, if $a\equiv 0$ $(\mod 2)$, then either Case (b-1) or
Case (b-3) happens and the deform-and-fill procedure applied to
$(S^3-K)\times_{\rho_1}X$ yields directly closed K3T $7$-manifolds.
If $a\equiv 1$ $(\mod 2)$, then Case (b-2) happens and the
deform-and-fill applied to $(S^3-K)\times_{\rho_1}X$ yields K3T
$7$-spaces with $A_1$-singularities. After blowups, this yields
also closed K3T $7$-manifolds.

\noindent\hspace{14cm} $\Box$

\bigskip

\noindent
{\bf Example 2.3.2
 [K3T with boundary ${\Bbb T}^2\times X$ via Whitehead link].}
Let $L=K_1\cup K_2$ be the {\it Whitehead link} in $S^3$,
as shown in {\sc Figure 2-3-2}.
\begin{figure}[htbp]
\setcaption{{\sc Figure 2-3-2.}
\baselineskip 14pt
 The Whitehead link in $S^3$.
} % end-setcaption
\centerline{\psfig{figure=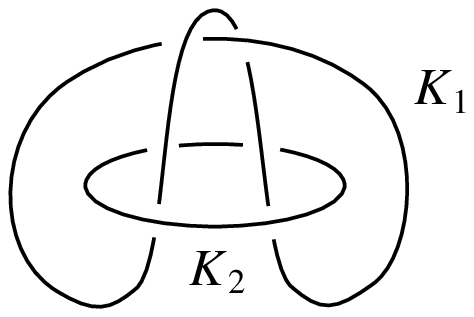,width=13cm,caption=}}
\end{figure}
Then $S^3-L$ is hyperbolic and
$H^1(S^3-L;{\Bbb Z}_2)={\Bbb Z}_1\oplus{\Bbb Z}_2$. From
Fact 2.1.1.1, $(S^3-K)_{(a_1,b_2;a_2,b_2)}$ is hyperbolic for
$a_1^2+b_1^2$, $a_2^2+b_2^2$ large enough.

The elements in $H^1(S^3-L;{\Bbb Z}_2)$ are
$$
 \rho_{\epsilon_1\epsilon_2}\; :\;
      m_i\;\longmapsto\;\iota^{\epsilon_i}\,, \hspace{1ex}
                 (\,l_i\;\longrightarrow\;\Id_X\,)\,,
                             \hspace{1em}\mbox{$i=1$, $2$}\,,
$$
where $\epsilon_1,\epsilon_2$ is $0$ or $1$. To obtain a K3T
$7$-manifold with boundary ${\Bbb T}^2\times X$, one considers
either $(S^3-L)\times_{\rho_{01}}X$ with $(a_1,b_1)=\infty$ or
$(S^3-L)\times_{\rho_{10}}X$ with $(a_2,b_2)=\infty$. In the
former case, if $a_2\equiv 0$ $(\mod 2)$, then either Case (b-1)
or Case (b-3) happens and the deform-and-fill procedure applied to
$(S^3-K)\times_{\rho_{01}}X$ yields directly a K3T $7$-manifold with
boundary ${\Bbb T}^2\times X$. If $a_2\equiv 1$ $(\mod 2)$, then
Case (b-2) happens and the deform-and-fill applied to
$(S^3-K)\times_{\rho_{01}}X$ yields a K3T $7$-space with
$A_1$-singularities. After blowups, this yields also a K3T
$7$-manifold with boundary ${\Bbb T}^2\times X$.
Similarly for the latter case.

\noindent\hspace{14cm} $\Box$

\bigskip

\noindent
{\bf Example 2.3.3 [K3T hierarchy].}
Let $K$ be the figure-8 knot in Example 2.3.1 and $L$ be the
Whitehead link in Example 2.3.2. With the same notations as in the
corresponding examples, $(S^3-K)_{(a,b)}$ is homeomorphic to
$(S^3-L)_{(a,b;1,-1)}$ ([Ro]), cf.\ {\sc Figure 2-3-3}. Consequently,
$(S^3-K)\times_{\rho_1}X$ is bundle-isomorphic to the bundle-filling
of $(S^3-L)\times_{\rho_{10}}X$ with the Dehn surgery coefficient
$(a,b;\infty)$. Any K3T $7$-manifold constructed from
$(S^3-K)\times_{\rho_1}X$ by deform-and-fill can also be constructed
from $(S^3-L)\times_{\rho_{10}}X$ by deform-and-fill. Thus the K3T
$7$-manifold $(S^3-K)\times_{\rho_1}X$ and its associated K3T family
are all descendants of $(S^3-L)\times_{\rho_{10}}X$.
\begin{figure}[htbp]
\setcaption{{\sc Figure 2-3-3.}
\baselineskip 14pt
 The figure-8 knot complement in $S^3$ is obtainable from the
 the Whitehead link complement in $S^3$ by the $(1,-1)$-surgery
 along $K_2$.
} % end-setcaption
\centerline{\psfig{figure=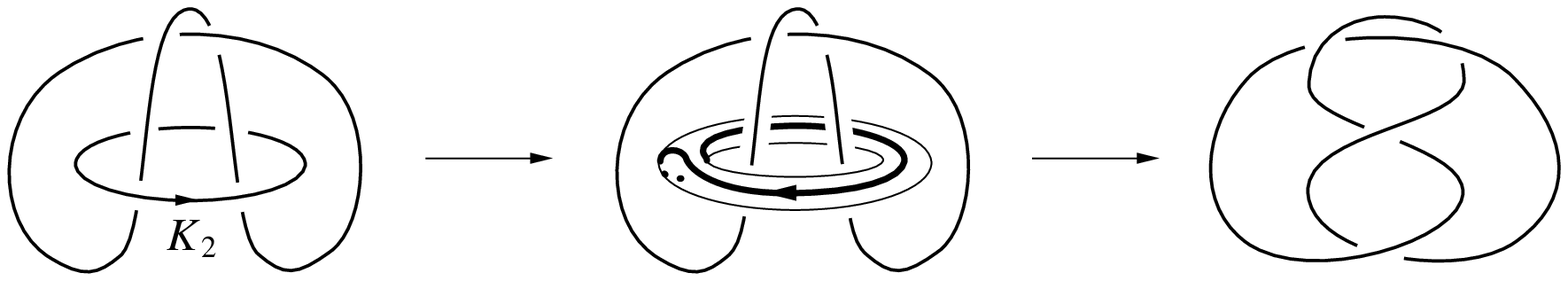,width=13cm,caption=}}
\end{figure}

\noindent\hspace{14cm} $\Box$

\bigskip

\noindent
{\bf Example 2.3.4 [general K3T in our class].}
Let $L=K_1\cup\,\cdots\,\cup K_i\cup\,\cdots\,\cup K_k$ be a link
in $S^3$ such that $S^3-L$ is hyperbolic. Examples of such links
from [Th1] are illustrated in {\sc Figure 2-3-4}. Let $(m_i,l_i)$
be the meridian and the longitute pair of $K_i$ and
$C=[\,c_{ij}\,]_{ij}=[\,\lk(K_i,K_j)\,]_{ij}$ be the linking matrix
of $L$ (with respect to some orientation of $L$ and $S^3$, which
does not enter the discussion after $\mod$ 2). Then the manifold
$(S^3-L)_{(a_1,b_1;\,\cdots\,; a_i,b_i;\,\cdots\,;a_k,b_k)}$
is hyperbolic if all $a_i^2+b_i^2$ are large.
\begin{figure}[htbp]
\setcaption{{\sc Figure 2-3-4.}
\baselineskip 14pt
 Some examples of hyperbolic links in $S^3$ from [Th1].
 For the notation, the link $C_n$ (resp.\ $D_{2n}$, $E_n$)
 has $n$ (resp.\ $2n$, $2n+3$) components.
} % end-setcaption
\centerline{\psfig{figure=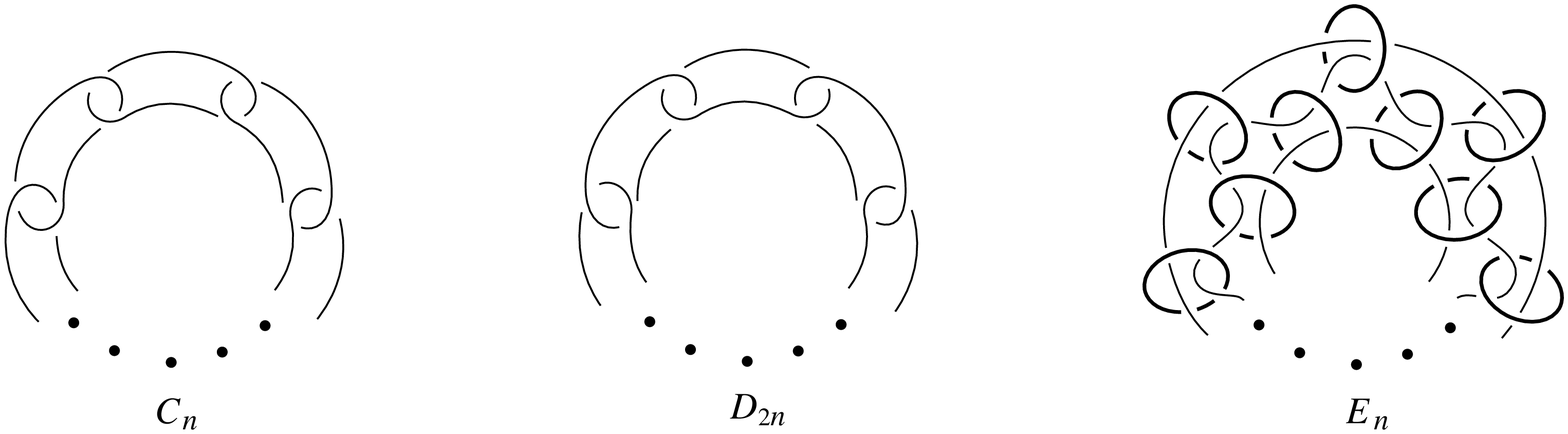,width=13cm,caption=}}
\end{figure}

The elements in $H^1(S^3-L;{\Bbb Z}_2)$ are given by 
$$
 \rho_{\epsilon_1\,\cdots\,\epsilon_k}\; :\;
 m_i\;\longmapsto\;\iota^{\epsilon_i}\,, \hspace{1ex}
 (\,l_i\;\longrightarrow\;
     \iota^{\mbox{\small $\Sigma$}_{j=1}^kc_{ij}\epsilon_j}\,)\,,
                            \hspace{1em}\mbox{$i=1,\,\ldots\,,k$}\,,
$$
where $\epsilon_1,\,\cdots\,,\epsilon_k$ are either $0$ or $1$.
Let $C_{(2)}$ be $C\;(\mod 2)$, ${\cal V}$ be the module
$\oplus_k\,{\Bbb Z}_2$, ${\cal V}_{i_1,\,\cdots,\,i_h}$,
$1\le i_1<\,\cdots\,<i_h\le k$, be the codimension $h$ submodule
that consists of points in ${\cal V}$ whose $i_s$-coordinate,
$s=1,\,\ldots\,h$ are $0$, and ${\Bbb T}_i$ be the $i$-th boundary
component of $S^3-\nu(L)$ associated to $K_i$. Then, the condition
that there is a nontrivial $\rho_{\epsilon_1,\,\cdots\,\epsilon_k}$
such that
$\rho_{\epsilon_1,\,\cdots\,\epsilon_k}(\pi_1({\Bbb T}_j))$
is trivial for $j\in\{i_1,\,\cdots\,i_h\}$ is that
${\cal V}_{i_1,\,\cdots,\,i_h}\cap
 C_{(2)}\,{\cal V}_{i_1,\,\cdots,\,i_h}\ne(0,\,\cdots\,,0)$.
For such $\rho_{\epsilon_1\,\cdots\,\epsilon_k}$, one can choose
$(a_i,b_i)$ to be $\infty$ for $i\in\{i_1,\,\cdots\,i_h\}$ and
some integer pair with large enough $a_i^2+b_i^2$ otherwise.
The corresponding deform-and-fill procedure applied to $\pi$
gives then a nontrivial K3T $7$-space
$$
 \pi^{\prime}\;:\; W^7 \;\longrightarrow\;
                     (S^3-L)_{(a_1,b_1;\,\cdots\,;a_k,b_k)}\,.
$$

Since 
$\rho_{\epsilon_1,\,\cdots\,\epsilon_k}(a_im_i+b_il_i)
 =\iota^{a_i\epsilon_i\,
  +\,b_i\,\mbox{\small$\Sigma$}_{j=1}^n\,c_{ij}\epsilon_j}$,
one has:
\begin{itemize}
 \item
 if $a_i\epsilon_i\,+\,b_i\sum_{j=1}^n c_{ij}\epsilon_j\equiv 0$
 $(\mod 2)$, then either Case (b-1) or Case (b-3) happens and it
 contributes no singularity to $W^7$; and

 \item
 if $a_i\epsilon_i\,+\,b_i\sum_{j=1}^n c_{ij}\epsilon_j\equiv 1$
 $(\mod 2)$, then Case (b-2) happens and this contributes
 $A_1$-singularities to $W^7$, which can be resolved by blowups.
\end{itemize}
In the end, this yields a K3T $7$-manifold
$$
 \widetilde{\pi}^{\prime}\;:\; {M^7}^{\prime}\;
  \longrightarrow\; (S^3-L)_{(a_1,b_1;\,\cdots\,;a_k,b_k)}
$$
with Calabi-Yau boundary
$\cup_h\,{\Bbb T}^2\hspace{-.4ex}\times\hspace{-.5ex} X$.

For the given three series of hyperbolic links $C_n$, $D_{2n}$,
and $E_n$, with respect to appropriate labelling of components
of $L$, their ${\Bbb Z}_2$ linking matrices $C_{(2)}$ are
respectively
$$
 \hspace{-3em}
 {\tiny
  \left[\,
   \begin{array}{cccccc}
    0       & 1 & 0      & \cdots & 0 & 1 \\
    1       & 0 & 1      &        &   & 0 \\
    0       & 1 & 0      &        &   &   \\
    \vdots  &   & \ddots & \ddots &   & \vdots \\
            &   &        &        & 1 & 0 \\
    0       &   & \cdots & 1      & 0 & 1 \\
    1       & 0 & \cdots & 0      & 1 & 0
   \end{array}\,
 \right]_{n\times n}
 }\,, % end-tiny
 {\tiny
  \left[\,
   \begin{array}{cccccc}
    0       & 1 & 0      & \cdots & 0 & 1 \\
    1       & 0 & 1      &        &   & 0 \\
    0       & 1 & 0      &        &   &   \\
    \vdots  &   & \ddots & \ddots &   & \vdots \\
            &   &        &        & 1 & 0 \\
    0       &   & \cdots & 1      & 0 & 1 \\
    1       & 0 & \cdots & 0      & 1 & 0
   \end{array}\,
 \right]_{2n\times 2n}
 }\,, % end-tiny
    \hspace{1em}\mbox{and}\hspace{1em}
 {\tiny
   \left[\,
    \mbox{
     \begin{tabular}{c|c}
      {\normalsize ${\rm O}_{3\times 3}$}
       & $\begin{array}{cccccc}
            1 & \cdots & 1 & 0 & \cdots & 0 \\
            1 & \cdots & 1 & 1 & \cdots & 1 \\
            0 & \cdots & 0 & 1 & \cdots & 1
          \end{array}$ \\ \hline \\[-1ex]
      $\begin{array}{ccc}
         1 & 1 & 0 \\
         \vdots & \vdots & \vdots \\
         1 & 1 & 0 \\
         0 & 1 & 1 \\
         \vdots & \vdots & \vdots \\
         0 & 1 & 1 
       \end{array}$   & {\normalsize ${\rm O}_{2n\times 2n}$}
     \end{tabular}
    }\, % end-mbox
   \right]
  }\,, % end-tiny
$$
where ${\rm O}_{s\times s}$ is the $s\times s$ zero-matrix.
By choosing $n$ large enough, one can guarantee the existence of
nontrivial ${\Bbb Z}_2$-solution to the system of homogeneous
equations on $(\epsilon_1,\,\cdots\,\epsilon_k)$:
$\epsilon_i \equiv \sum_{j=1}^k c_{ij}\epsilon_j \equiv 0$
$(\mod 2)$ with $i\in\{i_1,\,\cdots\,,i_h\}$. Following the previous
discussion, one then obtains nontrivial K3T $7$-manifolds with
boundary $\cup_h\,{\Bbb T}^2\hspace{-.4ex}\times\hspace{-.5ex}X$
for arbitrary $h$.

\noindent\hspace{14cm} $\Box$

\bigskip

\section{The deformation space of K3T 7-manifolds.}

Having in mind the potential application of K3T $7$-manifolds to
the compactification of M-theory, which involves $11$-dimensional
space-time, we study in this section the deformation space of the
K3T $7$-manifolds constructed in Sec.\ 2.
If the application of K3T to M-theory can be carried out solidly,
the deformation space of K3T should be related to the scalar fields
or their moduli space in the associated effective $4$-dimensional
theory.

\bigskip

\subsection{The deformation space of K3T 7-manifolds constructed.}
Recall the K3T $7$-manifold
$$
 \widetilde{\pi}^{\prime}\;:\;M^7\;\longrightarrow\;{M^3}^{\prime}\,.
$$
By construction, $\widetilde{\pi}^{\prime}$ admits a flat connection
over the complement of its set of critical values. The complex
structure of $X$ determines a complex structure on a generic fiber
of $\widetilde{\pi}^{\prime}$. The horizontal foliation by the flat
connection provides a transverse hyperbolic structure with respect
to $\widetilde{\pi}^{\prime}$, determined by the hyperbolic structure
of $N^3\subset{M^3}^{\prime}$. Deformations of the complex structures
on $X$ and the hyperbolic structures on $N^3$ induce deformations
of the fiber complex structures and the tranverse hyperbolic
structures of $\widetilde{\pi}^{\prime}$. Since these two
deformations are independent from each other, the deformation space
of the K3T $7$-manifolds constructed in Sec.\ 2 exhibits thus a
decomposition
$$
 \Def(\,\mbox{\rm K3T}\,)\;=\; \Def(X,\iota)\,
              \times\,\Def(N^3)\,\times{H^1(N^3,{\Bbb Z}_2)}\,,
$$
where the last component $H^1(N^3,{\Bbb Z}_2)$ corresponds to the
twistings by $\rho_X$.

It is known that $\Def(X,\iota)$ admits a K\"{a}hler structure.
And, as a corollary of works by many people, $\Def(N^3)$ admits
a special K\"{a}hler structure. In view of the goal that motivates
us the current work
- compactification of M-theory on a K3T $7$-manifold -,
the appearance of these structures is a very welcome feature.

Let us now discuss these structures in more detail.

\bigskip

\subsection{The K3 part of deformations of K3T.}

The deformation space for K3 surfaces has been studied intensively
by several authors. In this section, we summarize the results that
are related to the current work. Readers are referred to
[As], [Bo], [B-P-VV], [Dol], [G-W], [Ni4], [Ti], and [Vo]
for more details.

Let $L$ be the lattice $(-E_8)\oplus(-E_8)\oplus H\oplus H\oplus H$
of signature $(3,19)$ and $X$ be a K3 surface. Then the K3 lattice
$H^2(X;{\Bbb Z})$ with the cup product is isomorphic to $L$. The
choice of an isometry $\phi:H^2(X;{\Bbb Z})\rightarrow L$ determines
a point in $P(L_{\scriptsizeBbb C})$, corresponding to the complex
line $\phi_{\scriptsizeBbb C}(H^{2,0}(X;{\Bbb C}))$. This point is
called the period point of the marked K3 surface $(X,\phi)$ and the
set of all period points for a class of marked K3 surfaces is called
the period domain of that class. It gives the deformation space of
the complex structures of the K3 surfaces in that class.

Assume from now on that $X$ admits an antisymplectic involution
$\iota$, then $\iota$ induces an isometry $\iota^{\ast}$ on
$H^2(X;{\Bbb Z})$. Let $M$ be the Picard lattice
$H^{1,1}(X,{\Bbb R})\cap H^2(X;{\Bbb Z})$ of $X$ and
$T=M^{\perp}\subset H^2(X;{\Bbb Z})$ be the transcendental
lattice. Then $M$ coincides with $H^2(X;{\Bbb Z})^+$, the
fixed lattice of $\iota^{\ast}$, and $T$ coincides with
$H^2(X;{\Bbb Z})^-$, the sublattice, on which $\iota^{\ast}$ acts
by $-1$. Fix an embedding $i:M\rightarrow L$ and identify
$i(M)$ with $M$. Let
$$
 D_M\;=\;\{\,
 {\Bbb C}\Omega\in P(T\otimes{\Bbb C})\,|\,
  \Omega\wedge\Omega=0,\, \Omega\wedge\overline{\Omega}>0 \,\}
$$
and
$$
 \Delta_M\;=\;\{\,{\Bbb C}\Omega \in D_M\,|\,
    \mbox{there exists a nonzero $\alpha\in T$ such that
    $\alpha\cdot\Omega=0$}\,\}\,.
$$
Then ([Vo]) the period domain for the marked K3 surfaces $(X,\phi)$
with involution acting on $H^2(x;{\Bbb Z})$ as $i^{\ast}$ is given
by $D_M-\Delta_M$. A point in $\Delta_M$ corresponds to a K3 surface
$X^{\prime}$ with an involution $\iota^{\prime}$ whose induced
isometry ${\iota^{\prime}}^{\ast}$ on
$H^2(X^{\prime};{\Bbb Z})=H^2(X;{\Bbb Z})$ is different from
$\iota^{\ast}$. Presumably such $X^{\prime}$ has a larger Picard
lattice than $M$. 

Let us now turn to the geometry of the deformation space. By the
Hodge index theorem ([B-P-VV], [Dol]), $M$ has signature $(1,t)$
for some $t$ and, hence, $T$ has signature $(2,r)=(2,19-t)$.
A ${\Bbb C}\Omega\in D_M$ can be identified with the oriented
positive-definite $2$-plane in $T_{\scriptsizeBbb R}$ spanned by
$(\re\Omega,\im\Omega)$. This leads to the identification
$$
 D_M\; \cong\; O(2,r)/
  \hspace{-.1ex}\mbox{\raisebox{-.4ex}{$\SO(2)\times O(r)$}}\,.
$$
Since $O(2,r)$ has four components, $D_M$ has two isomorphic
components.

The inner product $Q$ on $T$ induces an $O(2,r)$-invariant
K\"{a}hler metric $ds^2$ on $D_M$ as follows. First $Q$ induces
a symmetric $2$-tensor, still denoted by $Q$, on
$T_{\scriptsizeBbb C}$. This defines a Hermitian inner product
of signature $(2,r)$ on $T_{\scriptsizeBbb C}$ by setting
$\langle v,w \rangle =Q(v,\overline{w})$. Let $\omega_0$ be
the associated K\"{a}hler form. Let
$(z_1,z_2,z_3,\,\cdots\,,z_{r+2})$ be the complex coordinates for
$(T_{\scriptsizeBbb C},\langle\,\cdot\,,\,\cdot\,\rangle)$ with
respect to an orthonormal basis. Let
$$
 K\;=\;\log Q(v,\overline{v})\;
  =\;\log(z_1\overline{z}_1 + z_2\overline{z}_2
   - z_3\overline{z}_3 -\, \cdots\, - z_{n+2}\overline{z}_{n+2})
$$
and 
$$
  \omega\;=\; \sqrt{-1}\:\partial\overline{\partial}K\;
   =\;\sqrt{-1}\;
    \frac{Q(v,\overline{v})\,\omega_0
         - Q(dv,\overline{v})\wedge Q(v,d\overline{v})
     }{Q(v,\overline{v})^2}\,,
$$
where $v=(z_1,\cdots,z_{r+2})$ and $dv=(dz_1,\cdots,dz_{r+2})$
in terms of the given coordinates. Note that $\omega$ is a
homogeneous $2$-form on $T_{\scriptsizeBbb C}-\{0\}$ and hence
descends to a $2$-form, also denoted by $\omega$, on
$P(T_{\scriptsizeBbb C})$. It is invariant under the induced
$O(2,r)$-action. Since this action is transitive on $D_M$, by
choosing a point in $D_M$, one can determine the signature of
the induced hermitian metric on $D_M$. For example, take
$p=[\,1:i:0:\,\cdots\,:0\,]\in D_M$; then, in terms of the
local coordinates
$$
 (\,w_0,\,w_1,\,\cdots,\,w_r\,)\;
 =\;(\,\mbox{\large $\frac{z_2}{z_1},\,
         \frac{z_3}{z_1},\,\cdots,\,\frac{z_{r+2}}{z_1}$}\,)\,,
$$
$D_M$ around $p$ is decribed by $1+w_0^2-w_1^2-\cdots-w_n^2=0$, and
the complex cotangent space $T^{\ast}_pD_M$ is given by $dw_0=0$.
Consequently,
$$
 \omega|_{T_pD_M}\;
 =\;- \mbox{\large $\frac{\sqrt{-1}}{2}$}\,
      \sum_{i=1}^r\, dw_i \wedge d\overline{w}_i\,.
$$
Thus $-\omega$ induces a positive-definite Hermitian metric $ds^2$
on $D_M$. This is the Weil-Petersson metric on $D_M$ ([Ti]). It has
negative holomorphic sectional curvatures that are bounded away
from zero ([Gr-S], [Ti]).

The above discussion also shows that $(D_M, ds^2)$ is complete as
a Riemannian manifold. Since $\Delta_M$ has complex codimension at
least $1$, we shall take $D_M$ as the deformation space for our
K3 surfaces.

\bigskip

\subsection{The Thurston part of deformations of K3T.}

The deformation space $\Def(M^3)$ of hyperbolic structures
associated to a complete hyperbolic $3$-manifold $M^3$ of finite
volume was explored by Thurston in [Th1]. Many details of his
method were later studied further in [N-Z] and [Yo1, Yo2]. With the
anticipation of relating $\Def(M^3)$ to some $4$-dimensional $N=2$
supersymmetric nonlinear $\sigma$-model, we shall rephrase their
results in terms of the complex symplectic language. A corollary
from their work is that there is a natural special K\"{a}hler
structure in a neighborhood of the complete $M^3\in\Def(M^3)$. The
associated integrable system and Seiberg-Witten-like $1$-form can
also be constructed, following [Fr1] and [D-M1]. We shall now
explain this in some details.

\bigskip

\begin{flushleft}
{\bf $\Def(M^3)$ from the complex symplectic viewpoint.}
\end{flushleft}

\begin{flushleft}
{\it (a) An isotropic embedding $\chi$ of $\Def(M^3)$ in
         $({\Bbb C}^{2n},\omega)$.}
\end{flushleft}
Recall from Sec.\ 1 the upper half-plane
${\Bbb H}_+
 =\{z^{\prime}\in{\Bbb C}\,|\,\im(z^{\prime})>0\}$
and the ideal tetrahedron $\Delta(z)$ of modulus $z\in{\Bbb H}_+$.
Let $M^3$ be a complete hyperbolic $3$-manifold of finite volume and 
$$
 M^3\;=\;\Delta(z_1^0)\,\cup\,\cdots\,\cup\,\Delta(z_n^0)
$$
be an ideal triangulation of $M^3$. Since the Euler characteristic
of $M^3$ is zero, the number of edges in the triangulation is the
same as the number $n$ of the tetrahedra. Deformations of the ideal
tetrahedra lead then to deformations of the hyperbolic structures
on $M^3$. Requiring the geometric angle around each edge of the
triangulation from the piecewise hyperbolic geometry be exactly
$2\pi$ leads to a system of rational constraint equations:
$$
 \prod_{\nu=1}^n\,
 (z_{\nu})^{r^{\prime}_{j\nu}}
                         (1-z_{\nu})^{r^{\prime\prime}_{j\nu}}\;
 =\; \prod_{\nu=1}^n\, (z_{\nu}^0)^{r^{\prime}_{j\nu}}
  (1-z_{\nu}^0)^{r^{\prime\prime}_{j\nu}}\;
                         (\,=\;\pm e^{2\pi i}\,)\;,
 \hspace{1em} j\;=\; 1\,,\,\ldots\,,\,n\;,
$$
where $z_i\in {\Bbb H}_+$ and the product in each equation is taken
in the universal covering group $\widetilde{{\Bbb C}^{\times}}$ of
${\Bbb C}^{\times}=({\Bbb C}-\{0\},\times)$.
There are some additional constraint equations coming from the
meridian to simple closed geodesics, around which the ends of ideal
tetrahedra wind. These equations are of the same form as the
constraint equations from edges. All together, $j$ runs from $1$ to
some $n^{\prime}\ge n$. The system of equations define
$\mbox{\it Def}\,(M^3)$ as an affine variety in
$({\Bbb H}_+)^n\,\subset\,{\Bbb C}^n$. This variety has a distinguished
point $[M^3_0]$ corresponding to the complete $M^3$. It is known from
[N-Z] and [Th1] that the complex dimension of $\Def(M^3)$ around
$[M^3_0]$ coincides with the number $h$ of cusps of $M^3$.

Let ${\Bbb C}^{2n}=({\Bbb C}^{2n}, \omega)$ be the complex vector
space with complex coordinates $(w_1,\,\cdots\,w_{2n})$ and
the complex symplectic form
$\omega\,=\,\Sigma_{i=1}^n\,dw_i\,\wedge\,dw_{n+i}$.
Consider the following holomorphic inclusion map
$$
\begin{array}{ccccc}
 \Theta & : & ({\Bbb H}_+)^n & \longrightarrow & {\Bbb C}^{2n} \\[1ex]
        &   & (z_1\,,\,\cdots\,,\,z_n) & \longmapsto
   & (\,\log\left(\frac{z_1}{z_1^0}\right)\,,\,
       \cdots\,,\,\log\left(\frac{z_n}{z_n^0}\right)\,;\,
      \log\left(\frac{1-z_1}{1-z_1^0}\right)\,,\,
       \cdots,\,\log\left(\frac{1-z_n}{1-z_n^0}\right)\,)\;,
\end{array}
$$
where we require that $\im(\log z) \in (0,\pi)$ and
$\im(\log\,(1-z)) \in (-\pi,0)$. Then the image $\image\Theta$ of
$\Theta$ is an embedded complex Lagrangian submanifold in
$({\Bbb C}^{2n},\omega_0)$.
Let $R$ be the matrix
$[\,R^{\prime}\,,R^{\prime\prime}\,]\;
 =\;[\,(r^{\prime}_{j\nu})\,,
      \,(r^{\prime\prime}_{j\nu})]_{n^{\prime}\times 2n}$
regarded as a linear map $\varphi_R$ from ${\Bbb C}^{2n}$
to ${\Bbb C}^{n^{\prime}}$ by matrix multiplication, then
$$
 \Def(M^3)\;
 =\; \image\Theta\,\cap\,\varphi_R^{-1}(0)\,.
$$
We shall denote this embedding of $\Def(M^3)$ in
$({\Bbb C}^{2n},\omega)$ by $\chi$. Note that $\chi$ send $[M^3_0]$
to the origin of ${\Bbb C}^{2n}$.

In this way, one realizes $\Def(M^3)$ as an isotropic analytic
variety in $({\Bbb C}^{2n},\omega)$ and one can choose a neighborhood
$U$ of $[M^3_0]\in\Def(M^3)$ that forms a complex $h$-dimensional
isotropic submanifold in $({\Bbb C}^{2n},\omega)$. In the following
discussion, we shall choose $U$ topologically a complex ball.

\bigskip

\begin{flushleft}
{\it (b) The tautological embedding $\psi$ of $U$ in
         $H^1(\partial M^3;{\Bbb C})$.}
\end{flushleft}
Recall from [Th1] (cf.\ Sec.\ 1) that the hyperbolic structure on
$M^3$ associated to $(z_1,\cdots, z_n)$ in $\Def(M^3)$ induces
a complex affine structure on $\partial M^3$ and hence determines
a holonomy representation from $H_1(\partial M^3;{\Bbb Z})$ to the
complex affine group $\Aff(1,{\Bbb C})$. The derivative of the
holonomy (i.e.\ ignoring the translation part of the honomomy)
gives then a holonomy map $\mu$ from $H_1(\partial M^3;{\Bbb Z})$
to ${\Bbb C}^{\times}$. Explicitly, given a class $[\gamma]$ in
$H_1(\partial M^3;{\Bbb Z})$ represented by a simplicial edge-loop
$\gamma$ with respect to the induced triangulation on $\partial M^3$
from the ideal triangulation of $M^3$, $\mu$ can be expressed in
terms of $z_\nu$'s as ([N-Z])
$$
\mu([\gamma])\; =\;
 \prod_{\nu=1}^n\,
  \left(\frac{z_{\nu}}{z_{\nu}^0}\right)^{c^{\prime}_{\nu}}
  \left(\frac{1-z_{\nu}}{1-z_{\nu}^0}
                       \right)^{c^{\prime\prime}_{\nu}}\,,
$$
where $c^{\prime}_{\nu}$ and $c^{\prime\prime}_{\nu}$ are some
integers and the factors appearing in the above product are from
the moduli of the triangle vertices touching $\gamma$ from the
right with respect to the orientation of $\gamma$ and
$\partial M^3$ ({\sc Figure 3-1}). This gives an embedding of the
neighborhood $U$ of $[M^3_0]$ from Part (a) into
$H^1(\partial M^3;{\Bbb C}^{\times})
 =\Hom(H_1(\partial M^3;{\Bbb Z}),{\Bbb C}^{\times})$
with $[M^3_0]$ mapped to the element $(\,\cdot\,\rightarrow 1)$ in
$H^1(\partial M^3;{\Bbb Z})$. Since $\log$ is a local embedding from
$({\Bbb C}^{\times}, 1)$ to $({\Bbb C},0)$, one can lift the image
of $U$ in $H^1(\partial M^3;{\Bbb C}^{\times})$ to
$H^1(\partial M^3;{\Bbb C})=\Hom(H_1(\partial M^3;{\Bbb Z}),{\Bbb C})$
by taking logarithm of the coefficient group.
\begin{figure}[htbp]
 \setcaption{{\sc Figure 3-1.}
  \baselineskip 14pt
  The triangle vertices that contribute to the holonomy of the
  induced complex affine structure along a loop $\gamma$
  (indicated by the thick line) on $\partial M^3$ are indicated by
  $\bullet\,$ and the orientation of the boundary ${\Bbb T}^2$ is
  indicated by $\circlearrowleft$. (Cf.\ {\sc Fig.\ 3} in [N-Z].)
 } % end-setcaption
 \centerline{\psfig{figure=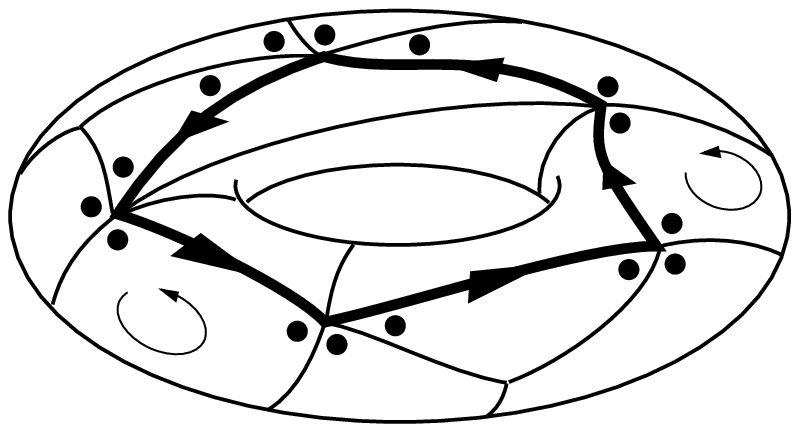,width=13cm,caption=}}
\end{figure}

We shall denote the latter embedding of $U$ by $\psi$. Note that
$[M^3_0]$ is mapped to the zero element of $H^1(M^3;{\Bbb C})$.

\bigskip

\begin{flushleft}
{\it (c) $\psi$ is Lagrangian.}
\end{flushleft}
First observe that, when tensored with ${\Bbb C}$, the cup product on
$H^1(\partial M^3;{\Bbb Z})$ induces a complex symplectic structure
$\omega^+$ on $H^1(\partial M^3;{\Bbb C})$ and the intersecting
pairing on $H_1(\partial M^3;{\Bbb Z})$ induces a complex symplectic
structure $\omega_-$ on $H_1(\partial M^3;{\Bbb C})$. 
The symplectic dual between $H^1(\partial M^3;{\Bbb C})$ and
$H_1(\partial M^3;{\Bbb C})$ coincides with the Poincar\'{e} dual
between the two.

Now let $[\gamma]\in H_1(\partial M^3;{\Bbb Z})$. Then $[\gamma]$
gives rise to a linear map $\varphi_{[\gamma]}$ from ${\Bbb C}^{2n}$
to ${\Bbb C}$ defined by
$$
 \varphi_{[\gamma]}(w_1,\,\cdots,\,w_n\,
                         ;\,w_{n+1},\,\cdots,\,w_{2n})\;
 =\; c^{\prime}_1w_1\,+\, \cdots c^{\prime}_nw_n\,
 +\,c^{\prime\prime}_1w_{n+1}\,+\,\cdots c^{\prime\prime}_nw_{2n}\,,
$$
where $c^{\prime}_\nu$, $c^{\prime\prime}_{\nu}$ are obtained as in
Part (b). The correspondence $[\gamma]\mapsto\varphi_{[\gamma]}$
gives a homomorphism from $H_1(\partial M^3;{\Bbb Z})$ to the dual
space                                                          
$({{\Bbb C}^{2n}}^{\ast},\omega^{\ast})$ of $({\Bbb C}^{2n},\omega)$.
This then induces a homomorphism
$\xi:H_1(\partial M^3;{\Bbb C})\rightarrow {{\Bbb C}^{2n}}^{\ast}$.
From Theorem 2.2 in [N-Z] and the discussion therearound, one has

\bigskip

\noindent
{\bf Fact 3.3.1.} ([N-Z]) {\it
 $\xi$ is injective and symplectic.
} % end-fact

\bigskip

\noindent
After taking the symplectic dual, one thus realizes
$H_1(\partial M^3;{\Bbb C})$ as a complex $2h$-dimensional symplectic
subspace ${\cal S}$ in $({\Bbb C}^{2n},\omega)$.

Recall the matrix $R$ from Part (a). Its row vectors span a complex
subspace in ${{\Bbb C}^{2n}}^{\ast}$. Let ${\cal R}$ be its symplectic
dual in ${\Bbb C}^{2n}$. Then Proposition 2.3 of [N-Z] can be
rephrased as

\bigskip

\noindent
{\bf Fact 3.3.2.} ([N-Z]) {\it
 ${\cal R}$ is isotropic and is contained in the symplectic
 orthogonal complement ${\cal S}^{\perp}$ of ${\cal S}$.
 Furthermore,
 $\varphi_R^{-1}(0)\,=\,{\cal R}\,\oplus\,{\cal S}$.
} % end-fact

\bigskip

One can now identify ${\cal S}$ with $H^1(\partial M^3;{\Bbb C})$
canonically since both are symplectic dual to
$H_1(\partial M^3;{\Bbb C})$. Let $\pi_{\cal S}$ be the projection
from $\varphi_R^{-1}(0)$ to ${\cal S}$. Then, by chasing the
definitions of the maps involved, one has

\bigskip

\noindent
{\bf Lemma 3.3.3.} {\it 
 $\pi_{\cal S}\circ\chi\,=\,\psi$.
} % end-lemma

\bigskip

\noindent
This shows that 

\bigskip

\noindent
{\bf Corollary 3.3.4.} {\it
 $\psi$ is a Lagrangian embedding of $U$ in
 $(H^1(\partial M^3;{\Bbb C}),\omega^+)$.
} % end-corollary

\bigskip

The relation of all the maps and spaces involved are indicated in
{\sc Figure 3-2} to make the above discussion more transparent.
\begin{figure}[htbp]
 \setcaption{{\sc Figure 3-2.}
 \baselineskip 14pt
  The relation of various maps and spaces.
 } % end-setcaption
 \centerline{\psfig{figure=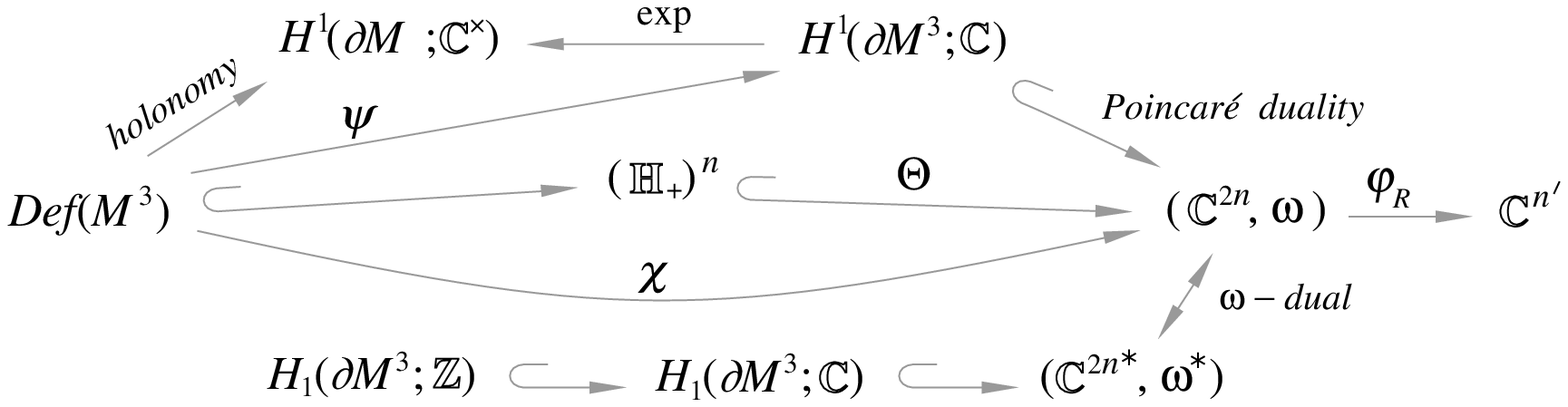,width=13cm,caption=}}
\end{figure}

\bigskip

\begin{flushleft}
{\bf The special K\"{a}hler structure on $U$ via $\psi$ and the
     associated integrable system.}
\end{flushleft}
With the preparation above, we can now spell out the
natural special K\"{a}hler structure on $U$.
Let $\partial M^3=\cup_h {\Bbb T}^2_i$ and $(\alpha_i,\beta_i)$
be a canonical basis of $H_1({\Bbb T}^2_i;{\Bbb Z})$ with
$\alpha_i\cdot\beta_i=1$. Then together
$(\alpha_1,\,\cdots\,,\alpha_h, \beta_1,\,\cdots\,,\beta_j)$
forms a canonical basis of $H_1(\partial M^3;{\Bbb Z})$. Let
$$
 {\cal A}^{\ast}=\Span_{\scriptsizeBbb C}(\alpha_1,\cdots,\alpha_h)
 \hspace{1em}\mbox{and}\hspace{1em}
 {\cal B}^{\ast}=\Span_{\scriptsizeBbb C}(\beta_1\cdots,\beta_h)
$$
be the isotropic framed subspaces in ${{\Bbb C}^{2n}}^{\ast}$ spanned
by $(\alpha_1,\,\cdots\,\alpha_h)$ and $(\beta_1,\,\cdots\,\beta_h)$
respectively and ${\cal A}$, ${\cal B}$ be their symplectic dual in
${\cal S}\subset {\Bbb C}^{2n}$. Note that $({\cal B},{\cal A})$
forms a transverse pair of framed Lagrangian subspaces in ${\cal S}$
and one can identify ${\cal S}$ as
$T^{\ast}{\cal B}={\cal B}\oplus(-{\cal A})$,
where $T^{\ast}{B}$ is the holomorphic cotangent bundle of ${\cal B}$
and $-{\cal A}$ is ${\cal A}$ with framing induced from
$(-\alpha_1,\,\cdots\,,-\alpha_h)$.

\bigskip

\noindent
{\it Remark 3.3.5.}
 As maps on ${\cal S}$, ${\cal A}^{\ast}$ corresponds the projection
 map $\pr_{\cal B}$ from ${\cal S}$ onto ${\cal B}$, while
 ${\cal B}^{\ast}$ corresponds to the projection map $\pr_{-{\cal A}}$
 from ${\cal S}$ onto $-{\cal A}$. This explains why 
 ${\cal S}=T^{\ast}{\cal B}={\cal B}\oplus(-{\cal A})$.
  
\bigskip

With $U$ identified with $\psi(U)$ in ${\cal S}$, $T_{[M^3_0]}U$,
the holomorphic tangent space of $U$ at $[M^3_0]$, is another
Lagrangian subspace in ${\cal S}$.

\bigskip

\noindent
{\bf Lemma 3.3.6 [transversality].} {\it
 $T_{[M^3_0]}U$ is transverse to both ${\cal B}$ and $-{\cal A}$.
} % end-lemma

\bigskip

\noindent
{\it Proof.}
Let $u_i$ be the element in ${{\Bbb C}^{2n}}^{\ast}$ associated to
$\alpha_i$ and $v_i$ be the element in ${{\Bbb C}^{2n}}^{\ast}$
associated to $\beta_i$. With notations from Remark 3.3.5,
$\pr_{\cal B}=(u_1,\,\cdots\,,u_h )$ and
$\pr_{-{\cal A}}=(v_1,\,\cdots\,,v_h)$. From [N-Z], when restricted
to $U\subset{\cal S}$, both $(u_1,\,\cdots\,,u_h )$ and
$(v_1,\,\cdots\,,v_h)$ form holomorphic coordinate charts around
$[M^3_0]\in U$ with the coordinates for $[M^3_0]$ being
$(0,\,\cdots\,,0)$. This concludes the lemma.

\noindent\hspace{14cm} $\Box$

\bigskip

Since $U$ is Lagrangian in ${\cal S}=T^{\ast}{\cal B}$, the
restriction to $U$ of the complex canonical $1$-form
$\alpha=\sum\,v_i\,du_i$ on $T^{\ast}{\cal B}$ vanishes. Thus 
$U$ can be realized as the graph of the holomorphic $1$-form
$d{\frak F}=\alpha|_U$ on ${\cal B}$ for some holomorphic function
${\frak F}$ on ${\cal B}$. Pulling back to $U$ via $\pr_{\cal B}$,
one can regard ${\frak F}$ as defined on $U$.
Lemma 3.3.6 implies that if one chooses the neighborhood $U$ of
$[M^3_0]\in\Def(M^3)$ appropriately, then $U$ is transverse to both
$\pr_{\cal B}$ and $\pr_{-{\cal A}}$. For such $U$, ${\frak F}$ then
serves as the prepotential that determines a special K\"{a}hler
structure $\omega_U$ on $U$
(cf.\ [Fr1]; also Theorem 2 and remarks in Sec.\ 4 of [Hi2]).
By specifying ${\frak F}(0,\,\cdots\,0)=0$, ${\frak F}$ is unique
on $U$. Let $K_U$ be the K\"{a}hler potential for $\omega_U$, then
in terms of the $u$-coordinates on $U$
$$
 v_i\;=\;\frac{{\partial{\frak F}}}{\partial u_i}\,,
  \hspace{3em}
 K_U\;=\;\mbox{$\frac{1}{2}$}\,\sum\,\im(v_i\,\overline{u_i})\;
  =\;\mbox{$\frac{1}{2}$}\,\sum
   \im(\frac{\partial {\frak F}}{\partial u_i}\,\overline{u_i})\,,
$$
and
$$
 \omega_U\;=\; \sqrt{-1}\,\partial\overline{\partial}\,K_U\;
  =\; \mbox{$\frac{\sqrt{-1}}{2}$}\,\sum\,
      \im(\frac{\partial{v_i}}{\partial u_j})\,
                                     du_i\wedge d\overline{u_j}\;
  =\; \mbox{$\frac{\sqrt{-1}}{2}$}\,\sum\,
      \im(\frac{\partial^2{\frak F}}{\partial u_i\,\partial u_j})\,
                                       du_i\wedge d\overline{u_j}\,.
$$
By Lemma 4.1 in [N-Z], at $u=(0,\,\cdots\,,0)\in U$ the Jacobian
matrix $[\partial v_i/\partial u_j]_{ij}$ is the diagonal matrix
$\Diag(\tau^0_1,\,\cdots\,,\tau^0_h)$, where $\tau^0_i$ is the
modulus of the complex structure on ${\Bbb T}^2_i$ associated to
the $i$-th cusp of $M^3$. Since every $\tau^0_i$ lies in the upper
half plane ${\Bbb H}_+$ and $U$ is connected, the K\"{a}hler
metric on $U$ associated to $\omega_U$ is positive-definite.

\bigskip

\noindent
{\it Remark 3.3.7 [independence of choices of $(\alpha_i,\beta_i)$].} 
In terms of the terminology in [Fr1], each
$({\cal A}^{\ast},{\cal B}^{\ast})$-pair provides a conjugate pair
of special holomorphic coordinate systems on $U$. Under different
choices of $(\alpha_i,\beta_i)$, the
$(u_1,\,\cdots\,,u_h, v_1,\,\cdots\,,v_h)$ transforms under
$\Pi_h\SL(2,{\Bbb Z})\subset\Sp(2h,{\Bbb Z})$, where the $i^{\,th}$
$\SL(2,{\Bbb Z})$ in the product acts only on $(u_i,v_i)$.
Thus the K\"{a}hler structure $\omega_U$ on $U$ constructed above
is independent of the choices of $(\alpha_i,\beta_i)$.
The twisted real part
$$
 (x_1,\,\cdots\,,x_h,y_1,\,\cdots\,y_h)\;
  =\;(\re u_1,\,\cdots\,,\re u_h, -\re v_1,\,\cdots\,,-\re v_h)
$$
of $(u_1,\,\cdots\,,u_h, v_1,\,\cdots\,,v_h)$ provides a real
Darboux coordinate system for $(U,\omega_U)$. Since the transition
functions among such real coordinate systems are also integral,
this defines a flat torsion free symplectic connection $\nabla$ on
$U$.

\bigskip

\noindent
{\it Remark 3.3.8 [naturality of ${\frak F}$].} ([N-Z] and [Yo1])
For $u=(u_1,\,\cdots\,u_h)\in U$ representing a hyperbolic
$3$-manifold $M^3_u$ obtained by a Dehn filling on some cusps of
$M^3$, the value ${\frak F}(u)$ is related to the volume
$\vol(M^3_u)$ and the Chern-Simons invariant $\CS(M^3_u)$ of $M^3_u$
by the identity
$$
 e^{\frac{2}{\pi}{\scriptsize\vol}(M^3_u)\,
                              +\,i\,{\scriptsize\CS}(M^3_u)}\:
   \Pi_i\,e^{{\scriptsize\length}(\gamma_i)\,
                      +\,i\,{\scriptsize\torsion}(\gamma_i)}\:   
   =\; e^{\frac{2}{\pi}{\scriptsize\vol}(M^3)\,
                               +\,i\,{\scriptsize\CS}(M^3)}\:
   e^{-\frac{1}{2\pi i}\,
   (-2\,+\,\sum\,u_i\,\frac{\partial}{\partial u_i})\,{\frak F}},
$$
where $\gamma_i$ are the simple geodesic loops assiciated to
the Dehn-filling and $\length(\gamma_i)$, $\torsion(\gamma_i)$
are the length and the torsion of $\gamma_i$
(cf.\ [Me], [N-Z], [Th2], and [Yo1]).
This indicates that ${\frak F}$ is a very natural holomorphic
function on $U$.

\bigskip

Now recall from [D-M2] and [Fr1] that

\bigskip

\noindent
{\bf Definition 3.3.9 [algebraic integrable system].} {\rm 
 An {\it algebraic integrable system} is a holomorphic map
 $\pi:Y\rightarrow M$, where $Y$ is a complex symplectic manifold
 with a holomorphic symplectic form $\eta$, such that
 \begin{quote}
  \hspace{-1.9em}(1)\hspace{.4ex}
   the fiber $Y_m=\pi^{-1}(m)$ for all $m\in M$ is a compact
   Lagrangian submanifols, hence an affine torus;

  \hspace{-1.9em}(2)\hspace{.4ex}
   there is a smooth family $[\rho]$ of cohomology classes
   $[\rho_m]\in H^{1,1}(Y_m)\cap H^2(Y_m;{\Bbb Z})$
   such that $[\rho_m]$ is a positive polarization of $Y_m$.
 \end{quote}
} % end-defintion

\bigskip

\noindent
Following previous discussions and notations, the holomorphic
$1$-forms $du_1$, $\cdots$, $du_h$, $dv_1$, $\cdots$, $dv_h$
generate a lattice $\Lambda$ of complex Lagrangian sections in
$T^{\ast}U$ that are flat with respect to $\nabla$.
Let $T^{\ast}U$ be the holomorphic cotangent bundle of $U$.
From Theorem 3.4 in [Fr1],
$Y=T^{\ast}U\hspace{-.4ex}/
   \hspace{-.2ex}\mbox{\raisebox{-.4ex}{$\Lambda$}}
     \rightarrow U$
is an algebraic integrable system over $U$ with $\eta$ from the
canonical holomorphic $2$-form on $T^{\ast}U$ and $\rho$ from the
dual of $\omega_U$. Since $\Lambda$ is independent of the choices
of $(\alpha_i,\beta_i)$ that defines $(u_i,v_i)$, $Y$ is
canonically associated to $U$.

\bigskip

\begin{flushleft}
{\bf The Seiberg-Witten-like 1-form $\lambda$ on $Y$.}
\end{flushleft}
Recall (e.g.\ [Don] and [D-M1]) that, for the integrable system
associated to the Seiberg-Witten theory of a gauge group or to
a complete family of Calabi-Yau threefolds, there is a natural
$1$-form $\lambda$ on the total space of the integrable system, 
whose periods along the fiber tori provide a conjugate pair of
holomorphic coordinate systems for the special K\"{a}hler
or the  projective special K\"{a}hler geometry on its base.
Something similar happens also to the integrable system
$(Y,\eta,\rho)$ over $U$.

\bigskip

\noindent
{\bf Lemma 3.3.10 [Seiberg-Witten-like 1-form].} {\it
 There exists a smooth $1$-form $\lambda$ on $Y$ such that, if 
 $(u_1,\,\cdots,\,u_h)$ and $(v_1,\,\cdots,\,v_h)$ are a conjugate
 pair of holomorphic coordinate systems on $U$ as discussed earlier
 and $C^{\prime}_1$, $\cdots$, $C^{\prime}_h$, $C^{\prime\prime}_1$,
 $\cdots,C^{\prime\prime}_h$ are the $1$-cycles on the fiber of $Y$
 associated to $du_1$, $\cdots$, $du_h$, $dv_1$, $\cdots$, $dv_h$
 respectively, then $\lambda$ satisfies the following properties:
 \begin{quote}
  \hspace{-1.9em}\parbox[t]{1.5em}{(1)}
  \parbox[t]{13.5cm}{
   the restriction of $\lambda$ to the fiber of $Y$ is closed
   and the period of $\lambda$ with respect to
   $(C^{\prime}_1,\cdots,C^{\prime}_h,
     C^{\prime\prime}_1,\cdots,C^{\prime\prime}_h)$
   is $(u_1,\cdots,u_h,v_1,\cdots,v_h)$;}

  \hspace{-1.9em}\parbox[t]{1.5em}{(2)}
  \parbox[t]{13.5cm}{
   $d\lambda=\eta$ and hence $\int_{C^{\prime}_i}\eta=du_i$ and
     $\int_{C^{\prime\prime}_i}\eta=dv_i$.}
 \end{quote}
} % end-lemma

\bigskip

\noindent
{\it Proof.}
Take $du_1, \cdots, du_h, dv_1,\cdots, dv_h$ as a real basis for
fibers of $T^{\ast}U$ and consider the following smooth
complex-valued function on $T^{\ast}U$
$$
\begin{array}{ccccc}
 f & : & T^{\ast}U & \longrightarrow
    & {\Bbb C}  \\[1ex]
   & & \sum\,c^{\prime}_i\,du_i\:
        +\: \sum\,c^{\prime\prime}_i\,dv_i & \longmapsto
    & \sum\,c^{\prime}_i\,u_i\:
       +\: \sum\,c^{\prime\prime}_i\,v_i\,.
\end{array}
$$
Notice that $f$ is independent of choices of the $(u,v)$-coordinates
used. Its differential
$$
 \mbox{
 $df\;=\;(\sum\,u_i\,dc^{\prime}_i\:
         +\: \sum\,v_i\,dc^{\prime\prime}_i)\,+\,
 (\sum\,c^{\prime}_i\,du_i\:
         +\: \sum\,c^{\prime\prime}_i\,dv_i)$ }
$$
is a complex-valued smooth $1$-form on $T^{\ast}U$. Let $t$ be
an element in the lattice $\Lambda$ of sections generated by
$du_1$, $\cdots$, $du_h$, $dv_1$, $\cdots$, $dv_h$ and ${\cal O}_t$
be the fiberwise translation of $T^{\ast}U$ generated by $t$, Then
$$
 {\cal O}_t^{\ast}\,df\,-\,df\;=\;t\,.
$$

Observe that the summand
$(\sum c^{\prime}_i du_i+\sum c^{\prime\prime}_i dv_i)$ of $df$
is exactly the canonical holomorphic $1$-form $\alpha$
on $T^{\ast}U$, which satisfies also
${\cal O}_t^{\ast}\alpha-\alpha=t$.
Consequently, the difference $df-\alpha$ is invariant under
${\cal O}_t$ and hence descends to a smooth complex-valued $1$-form
$\lambda$ on $Y$. That $\lambda$ satisfies both Property (1) and
Property (2) above follows by construction.
This concludes the proof.

\noindent\hspace{14cm} $\Box$

\bigskip

\noindent
{\it Remark 3.3.11 [relation to some natural $1$-forms supported on
                  the cusps of $M^3$].}
Recall from Sec.\ 2.1 the notations and the fact that each deformed
cusp ${\cal C}$ of $M^3$, if finite, is geometrically the quotient
$\overline{\widetilde{{\cal N}_{\varepsilon}}}/
  \hspace{-.1ex}\mbox{\raisebox{-.4ex}{$\widetilde{\mu}(
                                           \pi_1({\Bbb T}^2))$}}$.
The group of the induced $\widetilde{\mu}(\pi_1({\Bbb T}^2))$-action
on the ideal boundary
$\partial_{\infty}{\Bbb H}^3={\Bbb C}\cup{\infty}$
is a subgroup of the group of M\"{o}bius transformations that fix
$0$ and $\infty$. Since $d\log z=\frac{dz}{z}$ is invariant under
the latter group, one may first lift $\frac{dz}{z}$ to
$\widetilde{{\cal N}_{\varepsilon}}$ via the projection
$(z,t)\mapsto z$ and the covering map, and then project the
resulting $1$-form on $\widetilde{{\cal N}_{\varepsilon}}$ to a
complex-valued smooth $1$-form $\xi_{\cal C}$ on ${\cal C}$.
The discussion is similar for ${\cal C}$ infinite.
The periods of these $\xi_{\cal C}$ along a canonical basis
$(\alpha_1,\beta_1;\,\cdots\,;\alpha_h,\beta_h)$ for
$H_1(\partial M^3;{\Bbb Z})$ gives then a conjugate pair of
holomorphic coordinates $(u_1,\cdots,u_h)$ and $(v_1,\cdots,v_h)$
that appear in earlier discussions. This suggests another
natural flat bundle ${\cal T}$ over $U$ whose fiber ${\cal T}_u$
over $u\in U$ is the product of the toroidal components of
$\partial M^3$ with the complex affine structure determined by $u$.
From the product structure, one can lift $\xi_{\cal C}$ to
${\cal T}_u$ and sum them together to form a $1$-form along a fiber
of ${\cal T}$. Using the flat connection on ${\cal T}$, one obtains
then a smooth complex-valued $1$-form $\xi$ on ${\cal T}$.
By identifying ${\cal T}$ with the quotient
$(U\times H_1(\partial M^3;{\Bbb R}))/\hspace{-.1ex}
   \mbox{\raisebox{-.4ex}{$H_1(\partial M^3,{\Bbb Z})$}}$
and $H_1(\partial M^3;{\Bbb Z})$ with $\Lambda$, one has a bundle
isomorphism
$$
 \varphi\;:\;Y\;\longrightarrow\;{\cal T}\,.
$$
The pullback $1$-form $\varphi^{\ast}\xi$ is then cohomologous to
$\lambda$ constructed in Lemma 3.3.10.
In some way, $\varphi^{-1}$ resembles the Abel-Prym map in the
usual Seiberg-Witten theory (cf.\ [Don]).

\bigskip

\section{Remarks and examples on general K3T 7-manifolds.}

In this section, we discuss two ways in which the construction
in Sec.\ 2.2 can be generalized. This will provide us with many
other classes of K3T $7$-manifolds.

\bigskip

\begin{flushleft}
{\bf Flat K3T 7-manifolds from general $(X,\Aut(X))$.}
\end{flushleft}
The construction in Sec.\ 2.2 has an immediate generalization:
one may replace the K3 surface with involution by a K3 surface
$X$ with nontrivial group of automorphisms $\Aut(X)$. Here an
{\it automorphism} means a diffeomorphism that preserves the
complex structure.

Recall that every K3 surface is K\"{a}hler.
The {\it Strong Torelli Theorem} ([B-P-VV] and [L-P]) for K3
surfaces can be stated as:
{\it The group $\mbox{\it Aut}\,(X)$ of automorphisms of a K3
 surface $X$ coincides with the group of Hodge isometries of
 $H^2(X,{\Bbb Z})$ that preserve the K\"{a}hler cone of $X$.}
When $X$ is algebraic, let $S_X$ be the Picard lattice of $X$
and $W_X$ the group on $H^2(X;{\Bbb Z})$ generated by the
Picard-Lefschetz reflections $s_d:x\mapsto x+(x,d)d$ associated
to elements $d$ in $S_X$ with $(d,d)=-2$. Let
$\mbox{\it Isom}\,(S_X)$ be the group of isometries of $S_X$.
Then the quotient of $\mbox{\it Aut}\,(X)$ by the finite normal
subgroup that consists of automorphisms whose induced map on $S_X$
are trivial contains a subgroup of finite index that is isomorphic
to a subgroup of finite index in the quotient group
$\mbox{\it Isom}\,(S_X)/
      \hspace{-.1ex}\mbox{\raisebox{-.4ex}{$W_X$}}$
([PS-S]). These fundamental facts allow one to convert the study
of $\mbox{\it Aut}\,(X)$ to the study of lattices with a bilinear
form and the fundamental polyhedron of the $W_X$-action
(cf.\ [Ni1-5]).

Examples of K3 surfaces with an infinite automorphism group are
provided by exceptional K3 surfaces. These are algebraic K3
surfaces whose rank of $S_X$ equals the maximal possible number
$20$. They are all realizable as an elliptic pencil with
infinitely many sections ([S-I]). Once a section is fixed as the
identity section, the fiber then has an abelian group structure.
The fiberwise translation by a section gives then an automorphism
of $X$. This special subgroup of $\mbox{\it Aut}\,(X)$ is called
the {\it Mordell-Weil group} of $X$.

On the other hand, K3 surfaces with finite automorphism group
have been studied extensively by Nikulin ([Ni1-5]) and others.
Finite groups $G$ that can act on a K3 surface effectively and
leave the holomorphic $2$-form fixed were completely worked out
by Mukai ([Muk] and [Ni2]). He also gave the K3 surfaces on which
such $G$ act, as in {\sc Table 4-1}.
\begin{table}[htbp]
{\small
\begin{tabular}{|r|lrll|}  \hline
 $n^o$  & $G$        & order  & K3 surface
        & \raisebox{.6ex}{\rule{0em}{1em}} \\[.6ex]  \hline
  1     & $L_2(7)$   & $168$  & $x^3y+y^3z+z^3x+t^4=0$
                             in ${\smallBbb C}{\rm P}^3$
        & \raisebox{.6ex}{\rule{0em}{1em}} \\[2ex]
  2     & $\mbox{\it Alt}_6$
                     & $360$  & $\sum_{i=1}^6 x_i
                                 = \sum_{i=1}^6 x_i^2
                                 = \sum_{i=1}^6 x_i^3 =0$
                             in ${\smallBbb C}{\rm P}^5$ & \\[2ex]
  3     & $\mbox{\it Sym}_5$
                     & $120$  &  $\sum_{i=1}^5 x_i
                                  = \sum_{i=1}^6 x_i^2
                                  = \sum_{i=1}^5 x_i^3 =0$
                             in ${\smallBbb C}{\rm P}^5$ & \\[2ex]
  4     & $M_{20}=2^4\mbox{\it Alt}_5$
                     & $960$  & $x^4+y^4+z^4+t^4+12xyzt=0$
                            in ${\smallBbb C}{\rm P}^3$  & \\[2ex]
  5     & $F_{384}=4^2\mbox{\it Sym}_4$
                     & $384$  & $x^4+y^4+z^4+t^4=0$
                            in ${\smallBbb C}{\rm P}^3$  & \\[2ex]
  6     & $\mbox{\it Alt}_{4,4}=2^4\mbox{\it Alt}_{3,3}$
                     & $288$  & $\left\{\,\begin{array}{l}
                                        x^2+y^2+z^2=\sqrt{3}u^2\\
                                        x^2+\omega y^2
                                           +\omega^2 z^2
                                         =\sqrt{3}v^2 \\
                                        x^2+\omega^2 y^2
                                           +\omega z^2
                                            =\sqrt{3}w^2
                                          \end{array}\right.$
                          in ${\smallBbb C}{\rm P}^5$   & \\[5ex]
  7     & $T_{192}=(Q_8\ast Q_8)\rtimes\mbox{\it Sym}_3$
                     & $192$  & $x^4+y^4+z^4+t^4
                                 -2\sqrt{-3}(x^2y^2+z^2t^2)=0$
                           in ${\smallBbb C}{\rm P}^3$  & \\[2ex]
  8     & $H_{192}=2^4D_{12}$
                     & $192$  & $\left\{\,\begin{array}{l}
                                  x_1^2+x_3^2+x_5^2
                                        =x_2^2+x_4^2+x_6^2 \\
                                  x_1^2+x_4^2=x_2^2+x_5^2
                                        =x_3^2+x_6^2
                                  \end{array} \right. $
                            in ${\smallBbb C}{\rm P}^5$ & \\[4ex]
  9     & $N_{72}=3^2D_8$
                     & $72$   & $x_1^3+x_2^3+x_3^3+x_4^3
                                 =x_1x_2+x_3x_4+x_5^2=0$
                           in ${\smallBbb C}{\rm P}^4$  & \\[2ex]
 10     & $M_9=3^2Q_8$
                     & $72$   & double covering of
                                ${\smallBbb C}{\rm P}^2$
                                 branched over      & \\

                          & & & $x^6+y^6+z^6
                                 -10(x^3y^3+y^3z^3+z^3x^3)=0$
                                                    & \\[2ex]
 11     & $T_{48}=Q_8\rtimes\mbox{\it Sym}_3$
                     & $48$   & double covering of
                                ${\smallBbb C}{\rm P}^2$
                                  branched over  & \\
                          & & & $xy(x^4+y^4)+z^6=0$
                                             & \\[.6ex]  \hline
\end{tabular}
} % end-small
$$  % caption
\mbox{\parbox{11cm}{
 {\sc Table 4-1.} Mukai's table of K3 surfaces with finite
 symplectic automorphism group $G$. Readers are referred to
 [Muk] for the notation of various $G$ and more details.
 } }
$$  % end-caption
\end{table}
In particular, exactly the following fifteen groups can be
realized as such automorphism group for some K3 surface
([Ni2] and its Added-in-Proof):
\begin{quote}
{\small
 $({\smallBbb Z}_2)^k\,,\; 0\le k\le 4\,$;
 \hspace{1em}${\smallBbb Z}_3\,$; \hspace{1em}${\smallBbb Z}_4\,$;
 \hspace{1em}${\smallBbb Z}_5\,$; \hspace{1em}${\smallBbb Z}_6\,$;
 \hspace{1em}${\smallBbb Z}_7\,$; \hspace{1em}${\smallBbb Z}_8\,$;
 \hspace{1em}${\smallBbb Z}_2\oplus{\smallBbb Z}_4\,$;
 \hspace{1em}${\smallBbb Z}_2\oplus{\smallBbb Z}_6\,$;
 \hspace{1em}${\smallBbb Z}_3\oplus{\smallBbb Z}_3\,$;
 \hspace{1em}${\smallBbb Z}_4\oplus{\smallBbb Z}_4\,$.
} % end-small
\end{quote}
and the quotient orbifolds have $A$-singularities.

A program along this line, following the deform-and-fill procedure
in Sec.\ 2.2, involves the study of the following pieces:
\begin{enumerate}
 \item
 K3 surface $X$ with nontrivial $\Aut(X)$, how $\Aut(X)$ acts on
 $X$, and the deformation space of K3 surfaces that share isomorphic
 $\Aut(X)$. (Cf.\ {\sc Figure 4-1}.)
 \begin{figure}[htbp]
 \setcaption{{\sc Figure 4-1.}
  \baselineskip 14pt
  A schematic diagram for the deformation space of the complex
  structures of K3 surfaces with symmetry.
 } % end-setcaption
 \centerline{\psfig{figure=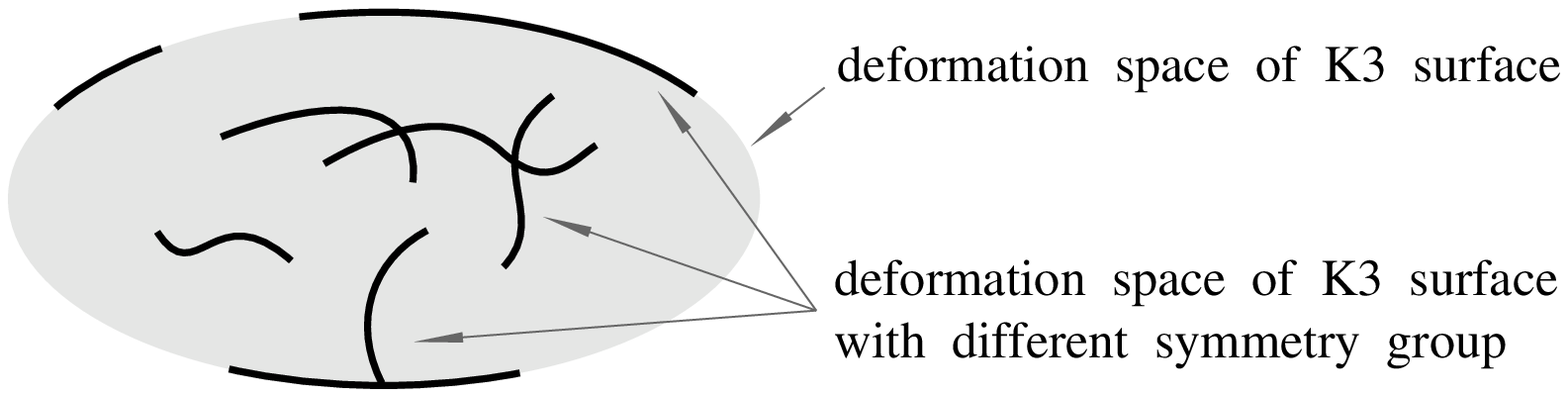,width=13cm,caption=}}
 \end{figure}

 \item
 Representations $\rho_X$ of $\pi_1(M^3)$ into $\Aut(X)$.
 (A computer code may be helpful here.
  Cf.\ e.g.\ [Fo], [GA], [Joh], [L-R], and [Ri1].)

 \item
 Singularities on $W^7$ obtained from $N^3\times_{\rho_X}X$ by the
 deform-and-fill procedure and their resolutions.

 \item
 Structures on the deformation space of the resulting K3T
 $7$-manifolds
 $$
  \Def(\,\mbox{\rm K3T}\,)\;
  =\; \Def(X,\Aut(X)) \times\Def(N^3) \times \Hom(\pi_1(N^3),\Aut(X))
 $$
 (cf.\ Sec.\ 3.1).
\end{enumerate}

With the information provided here, one can construct more classes
of K3T $7$-manifolds. The following example of K3-fibration, though
perhaps with nonhyperbolic base, is enough to illuminate the idea.

\bigskip

\noindent
{\bf Example 4.1 [K3-fibration via K3 with other symmetry].}
Let $X$ be the K3 surface given by the nonsingular complete
intersection of the quadric $x_1^2+x_2^2+x_3^2+x_4^2+x_5^2=0$ and
the cubic $x_1^3+x_2^3+x_3^3+x_4^3+x_5^3=0$ in ${\Bbb C}{\rm P}^4$
with homogeneous coordinates $[x_1:x_2:x_3:x_4:x_5]$. Then $X$
admits an action of the symmetric group $\mbox{\it Sym}\,_5$ of
degree $5$ induced from the action of $\mbox{\it Sym}\,_5$ on
${\Bbb C}{\rm P}^4$ by permutations of the homogeneous coordinates.
Let $K$ be the trefoil knot in $S^3$ ({\sc Figure 4-2}), whose
fundamental group $\pi_1(S^3-K)$ has a presentation
$\langle\,a,b\,|\,aba=bab\,\rangle$.
\begin{figure}[htbp]
\setcaption{{\sc Figure 4-2.}
\baselineskip 14pt
 The trefoil knot in $S^3$. (Its complement admits non-hyperbolic
 geometric structure.)
} % end-setcaption
\centerline{\psfig{figure=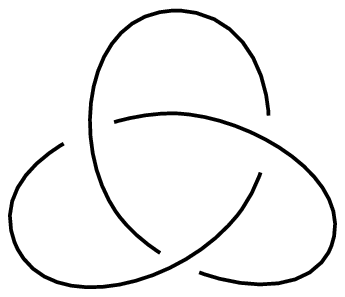,width=13cm,caption=}}
\end{figure}

Consider the representation $\rho$ from $\pi_1(S^3-K)$ to
$\mbox{\it Sym}\,_5$ generated by 
$$
 a\;\longmapsto\;(12345)\,,\hspace{3em} b\;\longmapsto\;(13542)
$$
([Fo], also [Ri1]; up to conjugation, this is the unique nonabelian
representation from $\pi_1(S^3-K)$ to $\mbox{\it Sym}\,_5$ that
sends the generators $a$ and $b$ to $5$-cycles).
Thus $\pi_1(S^3-K)$ acts on $X$ through $\rho$. Now let $W^7$ be
the quotient
$\widetilde{S^3-\nu(K)}\times X/
  \hspace{-.2ex}\mbox{\raisebox{-.4ex}{$\pi_1(S^3-K)$}}$,
where $\widetilde{S^3-\nu(K)}$ is the universal covering of
$S^3-\nu(K)$ and the group action is diagonal. Since the
$(1,1)$-loop along $K$ is sent by $\rho$ to the identity in
$\mbox{\it Sym}\,_5$, the boundary of this $7$-manifold is
diffeomorphic to $X_{\rho(a)}\times S^1$, where $X_{\rho(a)}$ is
the mapping torus
$X\times[0,1]/\hspace{-.2ex}
   \mbox{\raisebox{-.4ex}{$X\times\{0\}
                     \stackrel{\rho(a)}{\sim}X\times\{1\}$}}$.

To see how this boundary can be filled as in Sec.\ 2.2, notice that
$\rho(a)$ acts on $X$ with four fixed points of order $5$:
$$
 [1:\zeta:\zeta^2:\zeta^3:\zeta^4]\,,\hspace{1em}
 [1:\zeta^2:\zeta^4:\zeta:\zeta^3]\,,\hspace{1em}
 [1:\zeta^3:\zeta:\zeta^4:\zeta^2]\,,\hspace{1em}
 [1:\zeta^4:\zeta^3:\zeta^2:\zeta]\,,
$$
where $\zeta$ is a fifth root of $1$ and the quotient has $4$
$A_4$-singularities. Consequently, the action of the cyclic group
${\Bbb Z}_5=\langle\,f\,\rangle$ on $D^2\times X$ generated by
$$
 f\;:\; (z,x)\;\longmapsto \; (\zeta z\,,\,\rho(a)(x))
$$
has fixed points only on $\{O\}\times X$, which are exactly
the fixed points of $\rho(a)$ on $X$. By construction, the
quotient complex singular threefold has boundary $X_{\rho(a)}$.
To resolve the singularity, observe that the induced linear
map of $f$ on $T^{\prime}_p(D^2\times X)$, where $p$ is a fixed
point, can be put into the diagonal form
$\mbox{\it Diag}\,\{\zeta;\zeta,\zeta^{-1}\}$ if one chooses the
holomorphic coordinates of $X$ around $p$ appropriately. Hence,
if one blows up $D^2\times X$ at these fixed points and denotes the
new complex threefold by $\overline{Z}$, then $\langle\,f\,\rangle$
acts on $\overline{Z}$ freely. The quotient
$Z=\overline{Z}/
  \hspace{-.2ex}\mbox{\raisebox{-.4ex}{$\langle\,f\,\rangle$}}$
is smooth and fibred over
$D^2/
  \hspace{-.2ex}\mbox{\raisebox{-.4ex}{$\langle\,f\,\rangle$}}$,
which can be smoothened to a smooth $2$-disk again,
with central fiber
$\{X\,\sharp\,4\,(\overline{{\Bbb C}{\rm P}^2}\,
      \cup_{{\scriptsizeBbb C}{\rm P}^1}\,{\Bbb C}{\rm P}^2)\}/
  \hspace{-.2ex}\mbox{\raisebox{-.4ex}{$\langle\,f\,\rangle$}}$.
Consequently, $Z\times S^1$ fills $\partial W^7$.
Let $M^3$ be the $3$-manifold $(S^3-K)_{(1,1)}$.
After filled and smoothened, the resulting $M^7$ is K3-fibred over
$M^3$ with exceptional fiber
$\{X\,\sharp\,4\,(\overline{{\Bbb C}{\rm P}^2}\,
      \cup_{{\scriptsizeBbb C}{\rm P}^1}\,{\Bbb C}{\rm P}^2)\}/
  \hspace{-.2ex}\mbox{\raisebox{-.4ex}{$\langle\,f\,\rangle$}}$
over the core curve of the filling solid torus.

\noindent\hspace{14cm} $\Box$

\bigskip

\noindent
{\it Remark 4.2.}
Any K3T (or just K3-fibred) $7$-manifold thus constructed admits a
natural flat connection over the complement of the set of critical
values of the K3-fibration. Also its generic K3 fibers are all
biholomorphic. Thus, in some sense, it is in the category of
{\it flat analytic-K3T $7$-manifolds}, whose precise definition
will be left to the future after we gain better feeling about K3T.

\bigskip

\begin{flushleft}
{\bf From analytic-K3T to smooth-K3T.}
\end{flushleft}
In contrast with the flat analytic-K3T $7$-manifolds so far
discussed, let us give the reader some idea about general
smooth-K3T $7$-manifolds before ending.

Though no precise definition for a smooth-K3T $7$-manifold is
attempted in this paper, it is helpful to recall the following
definition in a well-understood case from [F-M2]:

\bigskip

\noindent 
{\bf Definition 4.3 [smooth-elliptic fibration].} {\rm
 A {\it $C^{\infty}$-elliptic surface} is a smooth map
 $\pi:S\rightarrow C$ from a closed, smooth, oriented $4$-manifold
 $S$ to a smooth oriented $2$-manifold $C$ such that for each point
 $p\in C$ there is an open disk $\Delta\subset C$ of $p$, a complex
 elliptic surface $Z\rightarrow\Delta$ and a smooth
 orientation-preserving diffeomorphism
 $\pi^{-1}(\Delta)\rightarrow Z$ such that the following diagram
 commutes:
 $$
  \begin{array}{cccl}
   \pi^{-1}(\Delta)
      & \stackrel{\cong}{\longrightarrow} & Z & \\
   \hspace{2ex}\downarrow \pi & & \downarrow & \\
   \Delta  & = & \Delta &.
  \end{array}
 $$
} % end-definition

\bigskip

\noindent
In other words, $S$ is the total space of pastings of local
elliptic fibrations $Z\rightarrow \Delta$ with smooth transition
functions. In the following, we shall use the word ``smooth" to
mean $C^r$ for some $r$.

For our situation, the base manifold of the K3-fibration has
real dimension $3$; and there is no theory of a real family of 
complex manifolds developed yet. So it is not clear what wild things
may arise if one follows the above definition to define smooth-K3T
$7$-manifolds. However, there are situations for which one can still
have control of the degenerate fiber: e.g.\ one may requires that
locally the fibration is diffeomorphic to a K3-fibration of the
form $\pi:Z\rightarrow\Delta\times[0,1]$ with $\pi$ restricted
to each $\Delta\times\{t\}$ the ususal K3-fibration from a
complex $1$-parameter family of K3 surfaces, or of the form
$\pi:Z\rightarrow \Delta^3$ that comes from the restriction to a
real $3$-disk of a K3-fibration associated to a complex
$2$-parameter family of K3 surfaces.

Confined to such, there are two kinds of complex $1$-parameter
degenerate K3 surfaces that have played roles in the
string literature (e.g.\ [As] and [A-M2]):
\begin{itemize}
 \item
 {\it Orbifold degenerations} ({\it A-D-E singularities}):
 ([B-P-VV] and [Di])
 When a K3 surface admits disjoint A-D-E chains of $(-2)$ curves,
 one may contract them through a complex $1$-parameter family via
 deformation of complex structures. The result is then a
 degenerate K3 surface with A-D-E singularities. A huge class of
 such examples are provided by elliptic K3 surfaces with a section.
 The Kodaira's table of singular elliptic fibers that can appear
 in an elliptic fibration of a complex surface contains various
 extended A-D-E Dynkin diagrams. For a K3 surface with a section,
 one can then contract the irreducible components of the singular
 fibers that are disjoint from the section and obtains a K3 surface
 with A-D-E singularities. From the work of Miranda and Persson
 ([M-P]), one knows that, while the total number of singular
 fibers of an elliptic K3 surface cannot exceed the Euler number
 $24$ of a K3 surface, there are {\it more than a thousand} of
 combinations of them that can truely happen. Via contraction,
 this provides then a big collection of degenerate K3 surfaces
 with various combinations A-D-E singularities. The nature of
 A-D-E singularities makes the resulting complex surface a
 complex orbifold.

 \item
 {\it Semistable degenerations}
 ({\it Kulikov degenerations of type II and III} ):
 ([F-S] and [Ku].)
 This is given by a family $\pi:X\rightarrow D^2$ of K3 surfaces
 over the unit disk $D^2=\{t\in{\Bbb C}\,|\,|t|\le1\}$ such that
 $X$ is smooth with trivial canonical bundle, all fibers
 $X_t=\pi^{-1}(t)$ are smooth K3 for $t\in D^2-\{0\}$, and the
 central fiber $X_0$ is a reduced divisor in $X$ with normal
 crossing singularities. There are only three situations:
 \begin{quote}
  \hspace{-2em}\parbox[t]{2.5em}{(I)}
  \parbox[t]{12cm}{$X_0$ is a smooth K3 surface.}

  \vspace{1ex}
  \hspace{-2em}\parbox[t]{2.5em}{(II)}
  \parbox[t]{12cm}{$X_0$ is a chain of elliptic ruled surfaces
   $V_1+\cdots+V_n$ whose associated simplicial complex of
   intersections is given by
   $$
    \begin{array}{lcl}
     \hspace{.6ex}\bullet
      \hspace{-.6ex}\rule[.5ex]{2em}{.2ex}\hspace{-1ex}\bullet
       \hspace{-.8ex}\rule[.5ex]{2em}{.2ex}\hspace{-.6ex}\bullet
       & \cdots
       & \bullet\hspace{-.6ex}\rule[.5ex]{2em}{.2ex}
          \hspace{-1ex}\bullet
           \hspace{-.8ex}\rule[.5ex]{2em}{.2ex}
            \hspace{-.6ex}\bullet \\[.4ex]
     V_1\hspace{1.2em}V_2 & \cdots
       & \hspace{1.8em}V_{n-1}\hspace{.4em}V_n
    \end{array}
   $$
   with rational surfaces on either end; the double curves
   $V_i\cap V_{i+1}$ that appear in the intersections are all
   elliptic curves.}

  \vspace{1ex}
  \hspace{-2em}\parbox[t]{2.5em}{(III)}
  \parbox[t]{12cm}{$X_0$ is a union of rational surfaces;
   the double curves on each irreducible component of $X_0$
   form a cycle of rational curves; and the associated simplicial
   complex of intersections is a triangulation of $S^2$.}
 \end{quote}
 (Note that the {\it simplicial complex of intersections}
  associated to a decomposition $X_0=\cup_i\,V_i$ by irreducible
  components is defined by assigning to each $V_i$ a vertex,
  to each non-empty $V_i\cap V_j$ an edge $e_{ij}$ connectiog
  $V_i$ and $V_j$, to each non-empty $V_i\cap V_j\cap V_k$ a face
  with boundary $e_{ij}\cup e_{jk}\cup e_{ki}$, etc.)
\end{itemize}

From Mumford's semistable reduction theorem ([K-K-M-SD]), an
orbifold - indeed any - degeneration $\pi:X\rightarrow D^2$ of     
algebraic K3 surfaces can always be converted into a semistable
degeneration $\pi^{\prime}:X^{\prime}\rightarrow {D^2}^{\prime}$
after blowups at the degenerate fiber $X_0=\pi^{-1}(0)$, a base
change ${D^2}^{\prime}\rightarrow D^2$ with
$t^{\prime}\mapsto t={t^{\prime}}^k$ for some positive integer
$k$, and then blowups again at the new central degenerate fiber of
$X\times_{D^2}{D^2}^{\prime}$ over ${D^2}^{\prime}$. When this
happens, the monodromy diffeomorphism $\tau^{\prime}$ of K3
associated to $\pi^{\prime}$ is the power $\tau^k$, where
$\tau$ is the monodromy diffeomorphism of K3 associated to $\pi$.

Combined with the construction in Sec.\ 2.2, this immediately
provides us with another big class of K3T $7$-manifolds.

\bigskip

\noindent
{\bf Example 4.4 [smooth-K3T].}
Let $E_i$, $i=1,\cdots,n$, be a disjoint collection of A-D-E chains
of embedded $2$-spheres of self-intersection $-2$ in a K3 surface
$X$ and $\tau_i$ be the associated monodromy diffeomorphism
supported in a small neighborhood of $E_i$. One can choose these
neighborhoods to be disjoint from each other so that $\tau_i$
commute with each other. Consequently, they generate an abelian
subgroup $\langle\,\tau_1,\,\cdots\,,\tau_n\,\rangle$ in
$\Diff(\mbox{\rm K3})$.

Let $L=K_1\cup\,\cdots\,\cup K_i\cup\,\cdots\,\cup K_k$ be a
hyperbolic link in $S^3$. As in Example 2.3.4, let $(m_i,l_i)$ be
the meridian and the longitute pair of $K_i$,
$C=[\,c_{ij}\,]_{ij}=[\,\lk(K_i,K_j)\,]_{ij}$ be the linking matrix
of $L$ with respect to some orientation of $L$ and $S^3$, and the
manifold from surgery
$(S^3-L)_{(a_1,b_1;\,\cdots\,; a_i,b_i;\,\cdots\,;a_k,b_k)}$
be hyperbolic. Since $\langle\,\tau_1,\,\cdots\,,\tau_n\,\rangle$
is abelian, a representation from $\pi_1(S^3-L)$ to
$\langle\,\tau_1,\,\cdots\,,\tau_n\,\rangle$
is determined by a representation $\rho_X$ from
$H_1(S^3-L;{\Bbb Z})$, which is a free abelian group generated by
$\{m_1,\,\cdots\,,m_k\}$,
to $\langle\,\tau_1,\,\cdots\,,\tau_n\,\rangle$. Let
$$
 \rho_X(m_i)\;
   =\;\tau_1^{\alpha_{i1}}\,\cdots\,\tau_n^{\alpha_{in}}\,,
$$
then
$$
 \rho_X(l_i)\;
  =\;\tau_1^{\mbox{\small $\Sigma$}_{j=1}^k c_{ij}\alpha_{j1}}\,
   \cdots\,\tau_n^{\mbox{\small $\Sigma$}_{j=1}^kc_{ij}\alpha_{jn}}
$$
and, for $(a_i,b_i)\ne\infty$,
$$
 \rho_X(a_im_i+b_il_i)\;
 =\; \tau_1^{a_i\alpha_{i1}
            +b_i\mbox{\small $\Sigma$}_{j=1}^k c_{ij}\alpha_{j1}}\,
     \cdots\,
     \tau_n^{a_i\alpha_{in}
            +b_i\mbox{\small $\Sigma$}_{j=1}^kc_{ij}\alpha_{jn}}\:.
$$
For such boundary component of $S^3-\nu(L)$, let us investigate
how the K3-bundle
$$
 \pi\;:\:(S^3-\nu(L))\times_{\rho_X} X\;
                              \longrightarrow\; S^3-\nu(L)\,.
$$
can be filled.

Consider the solid torus $V=\Delta\times S^1$, where $\Delta$ is a
compact disk in ${\Bbb C}$. Let $\hat{m}$ and $\hat{l}$ be a meridian
and a longitude of $V$ respectively. A representation $\rho$ from
$\pi_1(\partial V)=H_1(\partial V;{\Bbb Z})$ to
$\langle\,\tau_1,\,\cdots\,,\tau_n\,\rangle$ with
$$
 \rho(\hat{m})\;
  =\; f_{\hat{m}}\,=\,\tau_1^{k_1}\,\cdots\,\tau_n^{k_n}
   \hspace{2em}\mbox{and}\hspace{2em}
 \rho(\hat{l})\; =\; f_{\hat{l}}\,
  =\,\tau_1^{k_1^{\prime}}\,\cdots\,\tau_n^{k_n^{\prime}}\,,
$$
defines a flat K3-bundle:
$$
 \pr_1\;:\;
  \partial V\times_{\rho} X\,
    =\,({\Bbb R}^2\times X)/
       \hspace{-.2ex}\mbox{\raisebox{-.4ex}{$(\pi_1(\partial V),
                               \rho(\pi_1(\partial V)))$}}\;
                                   \longrightarrow\; \partial V\,,
$$
where the group action on ${\Bbb R}^2\times X$ is diagonal.
Let $\pr_2:Z\rightarrow \Delta$ be a $C^{\infty}$-K3 fibration
(cf.\ Definition 4.3) with degenerate fibers $X_{ij}$ over
$s_{ij}$ in the interior of $\Delta$ the singular K3 surface with
$E_i\subset X$ pinched to a point, where
$i\in \{i^{\prime}| k_{i^{\prime}} \ne 0 \}$ and
$j=1,\,\ldots\,,|k_i|$, such that the monodromy of the fibration
along $\partial\Delta$ (with induced orientation from that of
$\Delta$) is precisely $f_{\hat{m}}$.
Such fibration can be easily constructed by a fiber sum of the
corresponding complex $1$-parameter family degenerations
$\pi_{ij}:Z_{ij}\rightarrow (\Delta,0)$ that give each individual
isolated A-D-E singularity, as indicated in {\sc Figure 4-3}.
\begin{figure}[htbp]
 \setcaption{{\sc Figure 4-3.}
  \baselineskip 14pt
  A fiber sum of basic K3-fibrations associated to complex
  $1$-parameter family degenerations
  $\pi_{ij}:Z_{ij}\rightarrow (\Delta,0)$, realized as the fiber
  sum of each fibration to the trivial fibration
  $\Delta\times X\rightarrow \Delta$.
  The degenerate fiber of $\pi_{ij}$ is shown and 
  the occurrence of fiber sum is indicated by grey shade. 
 } % end-setcaption
 \centerline{\psfig{figure=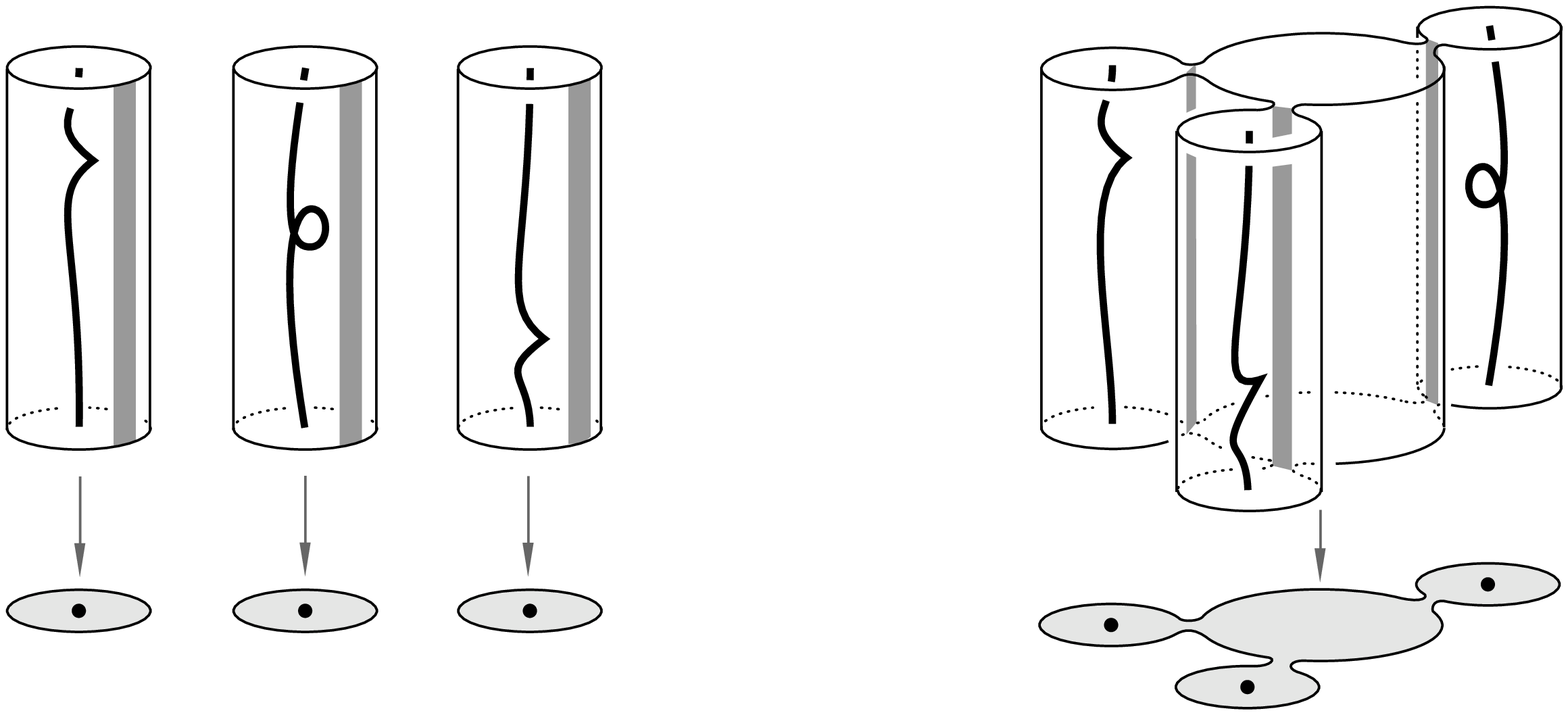,width=13cm,caption=}}
\end{figure}
Since $\tau_i$ can act on $Z_{ij}$ as a fibration $C^1$-automorphism
and $Z$ is constructed by fiber sum, $\tau_i$ acts also on $Z$.
Together with the fact that $\tau_1,\,\cdots\,,\tau_n$ commute,
this implies that $f_{\hat{l}}$ can act on $Z$ as a fibration
$C^1$-automorphism that projects to the identity map on $\Delta$
under $\pr_2$ and commutes with $f_{\hat{m}}$.

By taking the mapping torus $Z_{f_{\hat{l}}}$ of $Z$ with respect
to $f_{\hat{l}}$, one then obtains a K3-fibration
$$
 \pr\;:\; Z_{f_{\hat{l}}}\;\longrightarrow\; V\,,
$$
whose restriction to ${\Bbb T}^2=\partial V$ is equivalent to
$\pr_1$ and hence can be used to bundle-fill $\pr_1$. Applying
such bundle-filling to $\pi$ for each boundary ${\Bbb T}^2$ of
$S^3-\nu(L)$ with $(a_i.b_i)\ne\infty$, one obtains thus a
smooth-K3T $7$-manifold
$$
 \pi^{\prime}\;:\; M^7\;\longrightarrow\;
    (S^3-L)_{(a_1,b_1;\,\cdots\,; a_i,b_i;\,\cdots\,;a_k,b_k)}\,.
$$

This concludes the example.

\noindent\hspace{14cm} $\Box$

\bigskip

\noindent
{\it Remark 4.5 [flat smooth-K3T].}
Similar to the analytic category, one can also consider
$(X,\Diff(\mbox{\rm K3}))$ for a K3 surface $X$ and apply the
construction in Sec.\ 2.2 to obtain flat smooth-K3T $7$-manifolds.
A general theory of such involves in particular two issues:
\begin{quote}
 \hspace{-1.9em} (1)
  the structure of $\Diff(\mbox{\rm K3})$ and/or the mapping class
  group $\MCG(\mbox{\rm K3})$ of a K3 surface and

 \hspace{-1.9em} (2)
 representations of a $3$-manifold group $\pi_1(N^3)$ into
 $\Diff(\mbox{\rm K3})$ or $\MCG(\mbox{\rm K3})$,
\end{quote}
which by themselves are already very challenging.
As is well-known, the mapping class group of a compact Riemann
surface is finitely generated by the Dehn twists along a system
of simple loops (e.g.\ [Bi] and [H-T]); it is interesting to
know if something similar may happen for K3 surfaces.

\bigskip

\noindent
{\it Remark 4.6} However, in the smooth category, it is not clear
to us what kind of finite dimensional deformation space can be
associated to a K3T $7$-manifold.

\bigskip

\section{Issues on applications to M-theory compactification.}

K3T $7$-manifolds are constructed as natural interpolating
$7$-manifolds among some K3-fibred Calabi-Yau threefolds. This 
is meant to reflects the physical fact that M-theory interpolates
various string theories. In this section, we list some issues
for further study in order to understand the role of K3T
$7$-manifolds in M-theory.

\bigskip

\noindent $\bullet$
{\bf K3T and supergravity.}
M-theory has $11$-dimensional supergravity as its low-enery limit
([H-W1]). The field contents of the latter [C-J-S] consist of
 the $11$-bein $e^a_{\mu}$ (graviton),
 a Majorana spin-$\frac{3}{2}$ field $\psi_{\mu}$ (gravitino),
 and a $3$-form $A_{\mu\nu\rho}$ as gauge tensor.
In a local chart, the bosonic part of the field equations in the
theory reads
(cf.\ [Fre], summation convention for repeating indices assumed)
\begin{eqnarray*}
 R_{mn}-\frac{1}{2}g_{mn}R   & =
  & -\frac{1}{3}(H_{mpql}{H_n}^{pql}-\frac{1}{8}g_{mn}H^2) \\
 \nabla_m H^{mpql}  & =
  & -\frac{1}{576} \varepsilon^{m_1 \,\cdots\, m_8pql}
                       H_{m_1\,\cdots\,m_4}H_{m_5\,\cdots\,m_8}\,,
\end{eqnarray*} 
where $g_{mn}=e^p_me^q_n\eta_{pq}$ with
$\eta_{pq}=\Diag(-1,1,\,\cdots\,,1)$ the standard metric on the
$11$-dimensional Minkowskian space-time ${\Bbb R}^{1+10}$,
$R_{mn}$ and $R$ are the Ricci and scaler curvature of $g$,
$\nabla$ the covariant derivative associated to $g$, $H=dA$, and
$\varepsilon=\varepsilon_{m_1\,\cdots\,m_{11}}$ is the standard
volume-form on ${\Bbb R}^{1+10}$. The imposition of various ansatz
and the requirement of the residual supersymmetry when compactified
to $4$ dimensions lead to constraints on the underlying geometry,
for example, existence of Killing spinors (e.g.\ [Fre]) and many
exact solutions (e.g.\ [C-R-W], [D-N-P1], and [vN-W]) to the above
equations. Though related to Calabi-Yau threefolds by taking
boundary and to K3-fibred Joyce manifolds as a generalization via
$3$-dimensional geometry, unlike either kind of manifolds, general
K3T $7$-manifolds do not seem to satisfy these known constraints,
nor is it clear where there exists a K3T $7$-manifold that supports
a solution to the general field equations above. Perhaps one has to
look for other possible interfaces between K3T and M.

\bigskip

\noindent $\bullet$
{\bf Deformation space of K3T vs.\ deformation space of Calabi-Yau.}
One such candidate interface is via deformation space. This is
rooted at the fundamental theorem on hyperbolic $3$-manifolds
([Ber1, Ber2], [McM], and [Su]), which states roughly that,
given a hyperbolic $3$-manifold $M^3$, the Teichm\"{u}ller space
of complete hyperbolic structures on $M^3$ is isomorphic to the
Teichm\"{u}ller space of complex structures on its ideal boundary
$\partial_{\infty}M^3$ by sending a hyperbolic structure on $M^3$
to the conformal structure it induces on $\partial_{\infty} M^3$.
We already saw this kind of ``{\it boundary-dictate-interior}"
behavior in Sec.\ 3, where hyperbolic structures on $N^3$ is
controlled by the complex affine structures on its toroidal
$\partial N^3$. For analytic K3T $7$-manifolds with Calabi-Yau
boundary, this suggests a close relation between
$\Def(\mbox{\rm K3T})$ and $\Def(\mbox{\rm Calabi-Yau})$,
perhaps via fibration, embedding, or limit
(cf.\ [B-D-F-P-T-P-Z] and [Le]).

For the class of K3T $7$-manifolds constructed in Sec.\ 2, let
$(u_1,\,\cdots\,,u_h)$ and $(v_1,\,\,,v_h)$ be the conjugate
coordinates systems for $U\subset \Def(M^3)$ in Sec.\ 3, then an
immediate map relating $\Def(\mbox{\rm K3T})$ and
$\Def(\mbox{\rm Calabi-Yau})$ is given by
$$
 \begin{array}{cccl}
  \Def(\,\mbox{K3T}\,)  & \longrightarrow
   & \Def_c(\,\partial (\mbox{K3T})\,)
   & \subset\; \Def(\,\cup_{i=1}^h\:{\Bbb T}^2\times X\,) \\[1ex]
  (w\,;\,u_1,\,\cdots\,,u_h)  & \longmapsto
   &  \left(\, (\frac{v_1}{u_1},w)\,,\,
                 \cdots\,,\, (\frac{v_h}{u_h}, w)\,\right)\,, &
 \end{array}
$$
where $w$ is the complex structure of $X$ and
$\Def_c(\partial(\mbox{K3T}))$ is the deformation space of the
complex structures on the boundary $\cup_i\,{\Bbb T}^2\times X$.
Details of this map and what K\"{a}hler deformations of the
Calabi-Yau boundary are translated to on the K3T side remain to
be explored.

\bigskip

\noindent $\bullet$
{\bf K3T, dilogarithm, and CFT.}
The appearance of hyperbolic $3$-manifolds of finite volume as a
key ingredient in K3T $7$-manifolds has a mysterious feature.
We already saw in Sec.\ 3 that the deformation space of a such
$3$-manifold $M^3$ has a special K\"{a}hler structure with a
prepotential ${\frak F}$ related to the volume function $\vol$
(cf.\ Remark 3.3.8). It turns out that vol is related to the
dilogarithm function ([N-Z] and [Za]) defined by the analytic
continuation of the following power series
$$
 \Li_2(z)\;=\;\sum_{n=1}^{\infty}\,\frac{z^2}{n^2}
    \hspace{2em}\mbox{on}\hspace{2em}
              \{\,z\,|\,\|z\| <1\,\}\;\subset\; {\Bbb C}\,.
$$
Some role of this function in conformal field theory (CFT) was
explored in [Na] and [N-R-T]. Can one deepen their work and,
from which, find yet another way K3T and string/M-theory may
be related?

\vspace{3em}

We conclude this paper with these unresolved issues for future
pursuit.

%references
%endfootnotesize

\enddocument